\pdfoutput=1
\documentclass{article}

 \usepackage[main, final]{neurips_2026}

\usepackage[utf8]{inputenc} %
\usepackage[T1]{fontenc}    %
\usepackage{hyperref}       %
\usepackage{url}            %
\usepackage{graphicx}       %
\usepackage{float}          %
\usepackage{capt-of}        %
\usepackage{booktabs}       %
\usepackage{amsfonts}       %
\usepackage{nicefrac}       %
\usepackage{microtype}      %
\usepackage[table]{xcolor}  %
\usepackage{multirow}
\usepackage{amsmath}
\usepackage{amssymb}
\usepackage{mathtools}
\usepackage{pifont}

\title{Mental-Models for Multi-Agent Systems}

\author{%
  Hanan Gani \\
  \texttt{hgani@ucsd.edu} \\
  \And
  Lulu Shao \\
  \texttt{sishao@ucsd.edu} 
  \And
   Manmohan Chandraker \\
  \texttt{mkchandraker@ucsd.edu} \\
  \AND
  {\normalfont\normalsize University of California, San Diego} \\
  La Jolla, CA, 92093 \\
}

\begin{document}

\maketitle

\begin{figure*}[h]
  \centering
  \includegraphics[width=1.0\textwidth]{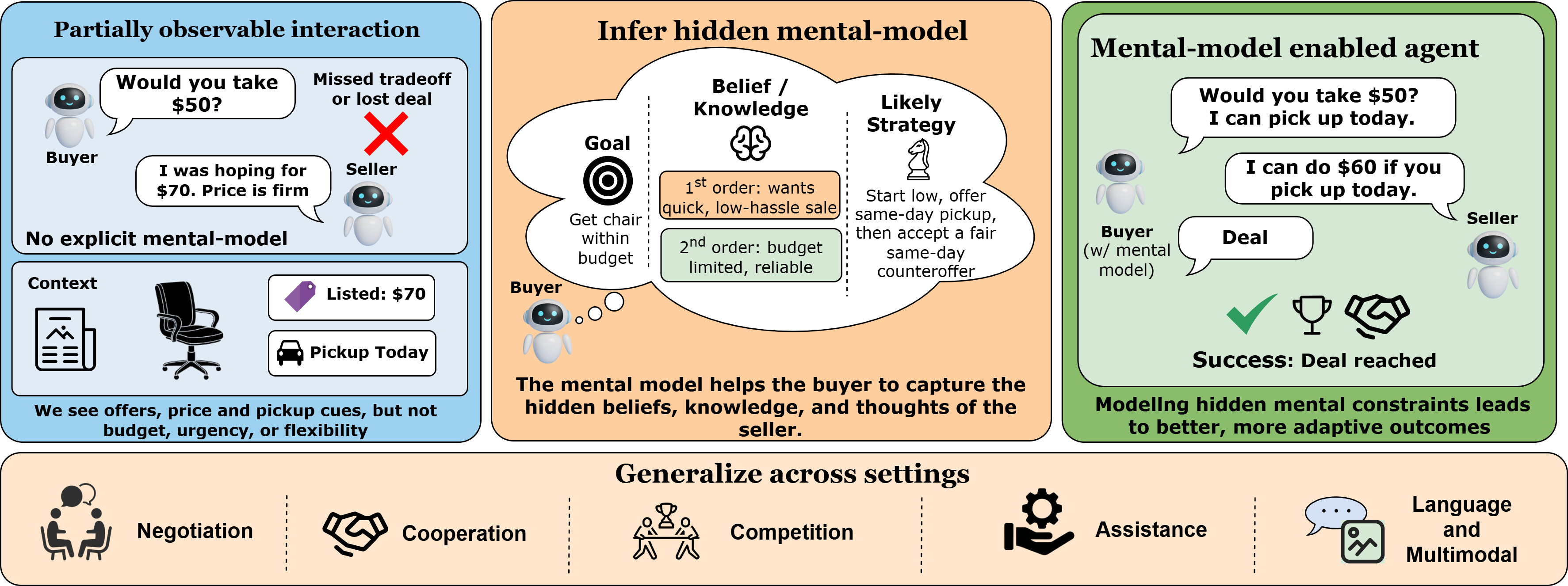}
  \vspace{-1em}
  \caption{\textbf{Mental-model-enabled decision making under partial observability.}
Observed behavior only partially reveals another agent's hidden goals, beliefs, constraints, and likely strategy. By inferring these latent mental states, a mental-model-enabled agent can choose more adaptive actions, improving outcomes while preserving cooperation across multi-agent settings.}
  \label{fig:teaser}
\end{figure*}

\begin{abstract}
Large foundation models have accelerated progress toward general-purpose agents that interact with humans and other agents through language and multimodal signals. However, robust multi-agent decision-making requires reasoning about what other agents know, intend, and are likely to do under partial observability. Current agentic systems often operate through prompt design, memory, or end-to-end behavioral shaping, but typically do not learn an explicit partner-state representation that can be reused as a decision variable across tasks. We introduce \emph{mental-model-enabled agents}, a framework that equips an agent with a latent mental model of its counterpart, allowing it to infer hidden beliefs, intentions, and likely reactions from the observed history and use these inferences to guide action selection. Our method learns an amortized recursive Theory-of-Mind representation, with first- and second-order mental-state structure, jointly with a belief-conditioned reward model that evaluates candidate actions relative to the inferred partner state. A policy is then learned under this belief-aware signal, yielding an agent that can act independently at inference time while retaining the benefits of explicit partner modeling. We evaluate the same framework on both language-only and multimodal benchmarks. Across these settings, explicit mental-state modeling consistently improves interaction quality and Theory-of-Mind performance over base agentic systems, showing that structured partner modeling is a useful inductive bias for general multi-agent systems. Our code is publicly available at \textcolor{magenta}{\url{https://github.com/hananshafi/Mental-Models}}
\end{abstract}

\section{Introduction}
\label{Introduction}
Modern AI, robotics, and cognitive science increasingly converge on a shared objective: building agents that can reason, plan, and act effectively in the world.
The core idea is that intelligent systems, both biological and artificial, must construct ``mental models'' of the world. This is in line with Kenneth Craik's seminal work \emph{The Nature of Explanation}, which proposed that organisms maintain small-scale internal representations of external reality~\citep{Craik1944TheNO}. Such mental models allow them to simulate possible actions, anticipate outcomes, and adapt effectively to changing
environments. However, in interactive environments, the world is partly constituted by other agents whose goals, beliefs, and strategies are not directly observable. As a result, successful action requires modeling not only the environment, but also other minds, a capacity commonly studied as Theory of Mind (ToM)~\citep{GopnikWellman1992}.

In humans, ToM is mediated by compact representations of a partner's latent mental states that support prediction and planning~\citep{GopnikWellman1992}. This perspective has motivated computational approaches that learn agent models directly from behavioral traces. ToMnet~\citep{pmlr-v80-rabinowitz18a}, for instance, showed that an observer can infer latent agent properties from past trajectories and use them to predict future behavior, demonstrating the value of explicit uncertainty over unobserved mental state and amortized inference that adapts rapidly to novel agents. These properties are particularly valuable in multi-agent settings, where partial observability is fundamental.

Large foundation models have accelerated progress toward general-purpose agents that interact with humans and other agents through language and multimodal signals. Yet robust interactive competence remains difficult. An agent must infer what other agents know, want, and are likely to do, and choose actions that improve outcomes under partial observability (Fig.\ref{fig:teaser}). Depending on the setting, this may involve cooperation, negotiation, assistance, competition, or adherence to conventions and norms.  Empirically, despite strong conversational ability, current LLMs lag behind humans on core ingredients of interactive reasoning, including Theory-of-Mind-style inference~\citep{SapEtAl2022NeuralToM,Ullman2023TrivialAlterationsToM,ShapiraEtAl2023CleverHans}, adherence to interaction norms~\citep{WeidingerEtAl2021EthicalSocialRisks}, and robust performance across goal-driven multi-agent scenarios~\citep{ZhouEtAl2024SOTOPIA}. Humans succeed in such settings because they rapidly form and update beliefs about others from limited evidence~\citep{KihlstromCantor2020SocialIntelligence,Tomasello2021BecomingHuman}. By contrast, current LLM agents typically do not maintain an explicit model of a partner's hidden goals, beliefs, or strategies. As a result, they may misread intent, commit to brittle plans, and struggle to adapt when the interaction unfolds in unexpected ways.

Most prior work on LLM-based multi-agent systems improves behavior through shaping or scaffolding rather than explicit partner-state estimation. Existing approaches include instruction tuning and preference-based reinforcement learning that steer generation~\citep{Ouyang2022InstructGPT,Shao2024DeepSeekR1,WangEtAl2024SOTOPIAPi}, prompt-level mechanisms such as self-critique, reflection, and memory that condition the agent on textual summaries of prior interaction~\citep{Bai2022ConstitutionalAI,Park2023GenerativeAgents}, and multi-turn synthetic self-play that induces complex interaction skills such as consensus-building~\citep{ni2025collaborative}. Across these approaches, partner dynamics are represented only implicitly either in the prompt text or trajectory summaries and not embedded in policy's latent dynamics. The result is the absence of an explicit belief state with clear semantics and quantified uncertainty, which limits interpretability, complicates reuse across tasks, and prevents partner modeling from being integrated systematically into different policy learners.

A long line of work in sequential decision-making under partial observability suggests that effective multi-agent behavior depends on maintaining beliefs over hidden state, including other agents' objectives, information states, and internal models~\citep{Kaelbling1998POMDP,DoshiGmytrasiewicz2011IPOMDP,DoshiZengChen2009GraphicalIPOMDP}. Extending this principle to foundation-model agents is nontrivial. When beliefs about a partner are expressed only as short natural-language hypotheses or serialized summaries appended to prompts, they remain limited in expressivity, and downstream action selection becomes tightly coupled to prompt formulation. This motivates the central question of our work: \textit{can an agent learn an amortized latent belief state from interaction history and use it as a stable training signal for multi-agent policy learning?}

In this paper, we address this question with \emph{mental-model-enabled agents} for multi-agent systems under partial observability. Our approach learns a compact latent belief state from interaction history that captures hidden variables of other agents, including goals, beliefs, knowledge, strategies, and higher-order expectations. We then learn a reward model that evaluates candidate actions relative to this inferred state, rather than only the observable context. The resulting belief-conditioned reward is used to train a policy under the learned mental model, while the final policy remains standalone at inference time. Together, these components define a general framework for ToM-guided decision making in multi-agent systems across diverse settings.

To validate the generality of our method, we evaluate it on both language-only and multimodal benchmarks for multi-agent systems. For language-only settings, we consider \textsc{SOTOPIA}~\citep{ZhouEtAl2024SOTOPIA}, \textsc{Craigslist-bargain} \cite{he2018decoupling},  \textsc{ToMi}~\citep{le-etal-2019-revisiting}, and \textsc{BigToM}~\citep{gandhi2023understanding}; for multimodal evaluation, we use \textsc{MMRole}~\citep{dai2025mmrole}. We also convert existing resources into ToM-supervised training data by augmenting them with mental-state annotations; Appendix Table~\ref{tab:dataset-comparison} summarizes the additional supervision relative to the original releases. Across these benchmarks, explicit mental-state modeling consistently improves downstream performance. For example, on \textsc{Sotopia}, our approach with Qwen2.5-7B achieves an average relative gain of $11.5\%$ over the base model across two partner settings, while on \textsc{BigToM} it improves paired TB/FB accuracy by $26.7$~pp on average across the six reported settings.

In summary, our contributions are:
\begin{itemize}
    \item We propose a general framework for ToM-guided multi-agent systems under partial observability, in which an agent learns an explicit latent belief state over other agents from observed history and uses it to support downstream decision making.

    \item We introduce a recursive mental-state architecture and a coupled mental-state/reward learning objective that make first- and second-order partner representations directly useful for evaluating candidate actions.

    \item We show how the learned mental signal can be distilled into standalone policies, so that explicit partner modeling improves training while deployment remains efficient.

    \item We evaluate the framework across language-only and multimodal benchmarks, and augment these resources with additional ToM supervision, utility signals, rationales, and hard negatives to support training and analysis.
\end{itemize}

\section{Related Work}
\label{related-work}

\textbf{Multi-agent systems.}
Recent work has explored LLM-based multi-agent systems in which multiple agents coordinate through role specialization, dialogue, memory, and tool use. CAMEL studies cooperative role-conditioned interactions in an agent society setting \citep{li2023camel}. MetaGPT, ChatDev, and AutoGen emphasize structured workflows, standardized operating procedures, and programmable conversation patterns for collaborative problem solving \citep{hong2023metagpt,qian-etal-2024-chatdev,wu2024autogen}. These systems improve performance mainly through external orchestration, such as role decomposition, protocol design, and tool integration. Our work is complementary: rather than modifying only the interaction protocol, we focus on internal partner modeling under partial observability by learning a latent belief state and using it to shape downstream control.

\textbf{Interactive environments and agent training.}
Another line of work develops environments and training pipelines for interactive agents. \textsc{Sotopia} \cite{ZhouEtAl2024SOTOPIA} provides open-ended, multi-turn interactions with private goals and multidimensional evaluation, and subsequent work studies imitation and self-reinforcement, utterance-level reward design, and scalable simulation in this setting \citep{ZhouEtAl2024SOTOPIA,WangEtAl2024SOTOPIAPi,yu2025sotopia_rl,zhou2025sotopia_s4}. In the multimodal setting, \textsc{MMRole} introduces a framework for developing and evaluating multimodal role-playing agents using persona-grounded images, dialogues, and reward-model-based evaluation \citep{dai2025mmrole}. These works provide important testbeds and training recipes for interactive agents, but they generally do not learn an explicit amortized partner model that is reused as a decision variable across benchmarks. Our approach instead learns a latent partner representation from interaction history, couples it with reward modeling, and distills the resulting signal into a policy.

\textbf{Theory of Mind and agent modeling.} ToM aims to infer or represent agents' latent mental states
(beliefs, goals, intentions) to predict future behavior and mental states. ToMnet~\citep{pmlr-v80-rabinowitz18a} learns latent models of agents from behavioral traces to predict future actions and agent-specific characteristics. In language-only settings, ToMi, BigToM, ToMBench, and OpenToM provide increasingly controlled or comprehensive evaluations of belief tracking and mental-state reasoning \citep{le-etal-2019-revisiting,gandhi2023understanding,chen-etal-2024-tombench,xu-etal-2024-opentom}. SymbolicToM shows that explicit belief tracking can substantially improve ToM reasoning at decoding time \citep{sclar-etal-2023-minding}. In multimodal settings, MMToM-QA and MuMA-ToM extend ToM evaluation beyond text, and MuMA-ToM further introduces LIMP, a model-based inverse multi-agent planning approach for multimodal mental-state inference \citep{jin2024mmtomqa,shi2025muma}. Relative to this literature, our contribution is not a new ToM benchmark or a task-specific inference procedure. Instead, we learn an amortized latent mental state from interaction history, couple it to utility prediction, and use the resulting signal to train policies for general multi-agent interaction across both language-only and multimodal benchmarks.

\section{Methodology}
\label{sec:method}

\begin{figure*}[t]
  \centering
  \includegraphics[width=1.0\textwidth]{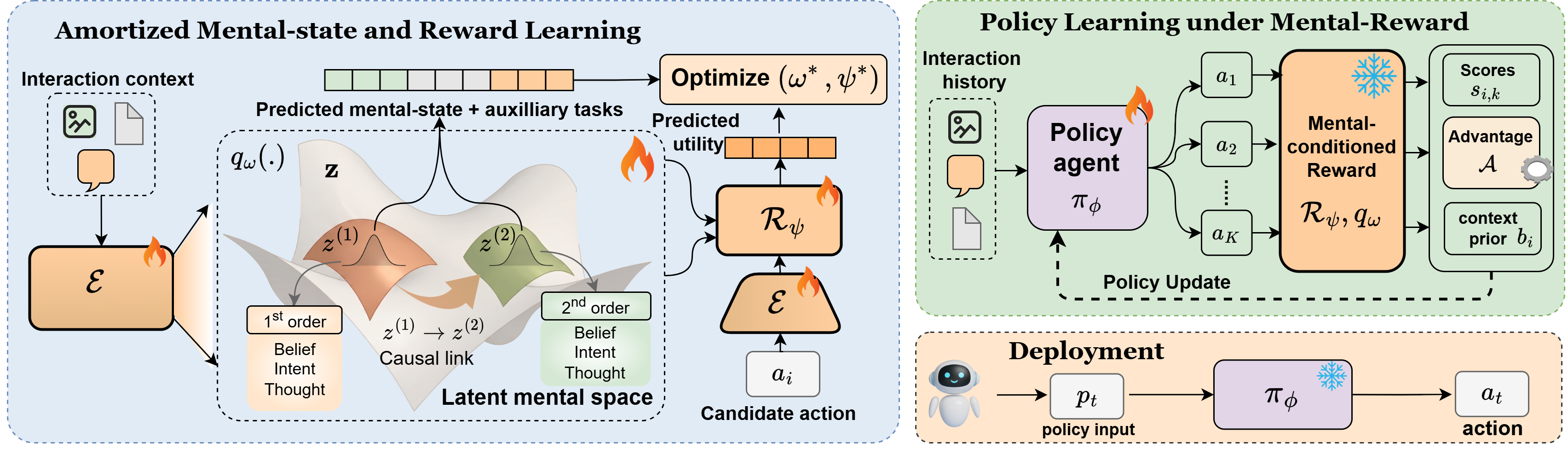}
  \vspace{-1em}
  \caption{\textbf{Method overview.} Phase 1 jointly learns an amortized recursive mental-model and a mental-conditioned reward model from multi-agent context. The latent contains first- and second-order belief, intent, and thought subspaces, which condition reward prediction. Phase 2 freezes the mental-state encoder and reward model, scores sampled policy actions with the learned mental reward, and optimizes the policy with mental-reward guidance. Phase 3 deploys only the trained policy in the multi-agent setting.}
  \label{fig:main-framework}
  \vspace{-1em}
\end{figure*}

We study multi-agent interaction under partial observability. As shown in Fig.~\ref{fig:main-framework}, we first learn an amortized latent mental-model of a counterpart's hidden state together with a utility model over candidate actions. We then use this learned signal to train a standalone policy. Finally, we deploy the policy directly at test time. We present the detailed mathematical formulation below.
For notational clarity, we present the two-agent case with one acting agent and one counterpart. 

\subsection{Amortized Mental-model Learning}
\label{sec:problem_formulation}

We consider sequential interaction over a finite horizon $T$. At time $t$, the acting agent receives an observation $o_t$, maintains the interaction history
$
h_t = (o_1, a_1, \dots, o_t),
$
and selects an action $a_t \in \mathcal{A}$. The action may be free-form language or a structured environment action. In multimodal settings, $o_t$ may also contain visual or other non-textual signals.

The effect of $a_t$ depends not only on the external environment, but also on latent mental properties of the counterpart, such as goals, beliefs, knowledge, capabilities, and strategies. We denote these hidden variables by $z_t$. A Bayes-optimal agent would maintain the posterior
$
p(z_t \mid h_t),
$
but exact inference is generally intractable in open-ended interaction. We therefore learn an amortized mental-state model
\begin{equation}
q_{\omega}(z_t \mid h_t) \approx p(z_t \mid h_t),
\label{eq:amortized_belief_main}
\end{equation}
which maps the observable history to a compact latent belief state.

However, ToM is inherently multi-order (as illustrated in Figure~\ref{fig:reward-vector-collapse}d). It requires reasoning not only about a partner's first-order state, but also about that partner's model of agent or environment. We capture this recursion by factorizing the mental-model latent as
\begin{equation}
q_{\omega}(z_t^{(1)},z_t^{(2)}\mid h_t)
=
q_{\omega_1}(z_t^{(1)}\mid h_t)\,
q_{\omega_2}(z_t^{(2)}\mid h_t,z_t^{(1)}).
\label{eq:recursive_factorization_main}
\end{equation}
Here $z_t^{(1)}$ represents the first-order partner state, while $z_t^{(2)}$ represents higher-order state conditioned on the first-order inference. Each mental-latent $z_t^{(\ell)}$ is decomposed into semantically meaningful subspaces representing belief, intent and thought-process, which are the fundamental elements of human theory-of-mind
\begin{equation}
z_t^{(\ell)}=
\big[
z_{t,\mathrm{belief}}^{(\ell)};
z_{t,\mathrm{intent}}^{(\ell)};
z_{t,\mathrm{thought}}^{(\ell)}
\big],
\qquad
\ell \in \{1,2\}
\label{eq:mental_subspaces}
\end{equation}
Using a single flat latent to represent both first- and second-order states can conflate two distinct estimation problems (Figure~\ref{fig:reward-vector-collapse}c,d). The recursive factorization keeps them connected without collapsing them. Practically, each latent factor is modeled as a multi-variate Gaussian distribution,
\begin{align}
q_{\omega_1}(z_t^{(1)} \mid h_t) &= \mathcal{N}\big(\mu_1(\mathcal{E}(h_t)), \operatorname{diag}(\sigma_1^2(\mathcal{E}(h_t)))\big) \\
q_{\omega_2}(z_t^{(2)} \mid h_t, z_t^{(1)}) &= \mathcal{N}\big(\mu_2([\mathcal{E}(h_t); z_t^{(1)}]), \operatorname{diag}(\sigma_2^2([\mathcal{E}(h_t); z_t^{(1)}]))\big)
\end{align}
Modeling latent factors as a Gaussian posterior family with a VAE-style parameterization gives the mental model a smooth geometry over beliefs, intents, and thoughts while representing uncertainty through the posterior variance. This lets the model represent a continuous distribution over plausible mental states rather than committing to a single point estimate, which is important because another agent's beliefs, intentions, and strategies are only partially observed and must be inferred under uncertainty. The first-order latent uses a standard normal prior, while the higher-order latent is regularized against a conditional Gaussian prior induced by $z_t^{(1)}$ so that the variational bottleneck preserves the recursive factorization. For source-specific mental-state supervision or proxy evidence $y_t$, the latent is learned with a compressed inference objective
\begin{equation}
\omega^\star
\in
\arg\min_{\omega}
\mathbb{E}_{(h_t,y_t)\sim\mathcal{D}}
\left[
\mathcal{D}_{\mathrm{tom}}\!\left(q_{\omega}(z_t\mid h_t),\, y_t\right)
+
\beta\,\Omega_{\mathrm{comp}}\!\left(q_{\omega}(z_t\mid h_t)\right)
\right],
\label{eq:belief_bottleneck_main}
\end{equation}

where the discrepancy term $\mathcal{D}_{\mathrm{tom}}$ grounds the latent in available mental-state evidence. The compression penalty $\Omega_{\mathrm{comp}}$ limits the capacity of the inferred latent state. In our implementation, it is instantiated as a recursive variational bottleneck with a Gaussian posterior, a history-independent prior for $z_t^{(1)}$, and a conditional prior for $z_t^{(2)}$.

\subsection{Mental-Model-Aware Utility}

A latent mental-model on its own cannot act as a decision variable and therefore cannot inform the agent what to do. To translate mental-model into action, we need a way to evaluate candidate actions against the inferred counterpart state. We therefore learn a reward utility function that scores arbitrary candidates conditioned on the mental-model latent. 

For an evaluation setting $e$, let
\(
U_e(h_t,a_t,\theta_t)\in\mathbb{R}^{D_e}
\)
be the outcome vector for action $a_t$ under the hidden counterpart state, and let $w_e\in\Delta^{D_e-1}$ be the scalarization of those dimensions. For example, an open-ended social setting may score goal completion and social quality, while a single-step ToM setting may score answer correctness. We learn a mental-model-conditioned utility predictor
\begin{equation}
\bar{\mathbf{u}}_{\omega,\psi}(h_t,a_t)
=
\mathbb{E}_{z_t\sim q_{\omega}(\cdot\mid h_t)}
\left[
\hat{\mathbf{u}}_{\psi}(h_t,a_t,z_t)
\right].
\label{eq:belief_aware_score_main}
\end{equation}
where  $\hat{\mathbf{u}}_{\psi}(h_t,a_t,z_t)$ is the outcome from the utility head $\mathcal{R}_{\psi}$.
This expectation turns the latent mental model into a score over candidate actions. It asks how good action $a_t$ is on average across the mental states that remain plausible given the history. The supervision used to fit $\mathcal{R}_\psi$ varies across benchmarks. Refer to Appendix \ref{app:detailed_method} for a detailed explanation.

Learning the mental-model and the utility separately tends to produce two complementary failure modes. A mental-model trained only on mental-state annotations becomes interpretable but decision-irrelevant. It captures what is describable about the partner rather than what is actionable. A reward model trained without pressure to use that latent explains its scores through surface features of the candidate response and ignores partner state entirely. We therefore optimize the two jointly, so that gradients from the reward objective shape the mental-model toward decision-relevance, and the mental-model in turn forces the reward model to route through it.
\begin{equation}
(\omega^{\star},\psi^{\star})
\in
\arg\min_{\omega,\psi}
\mathbb{E}_{x\sim\mathcal{D}}
\left[
\underbrace{\mathcal{J}_{\mathrm{rew}}(x;\omega,\psi)}_{\text{candidate reward}}
+
\underbrace{\mathcal{J}_{\mathrm{mm}}(x;\omega)}_{\text{mental-state inference}}
+
\underbrace{\mathcal{J}_{\mathrm{aux}}(x;\omega,\psi)}_{\text{auxiliaries}}
\right].
\label{eq:joint_objective}
\end{equation}
$\mathcal{J}_{\mathrm{rew}}$ fits the candidate-action scorer in Eq.~\eqref{eq:belief_aware_score_main}, including hard-negative ranking when contrastive responses are available. $\mathcal{J}_{\mathrm{mm}}$ anchors $z_t$ to first- and second-order beliefs, intents, thoughts, or other mental-state evidence through reconstruction. $\mathcal{J}_{\mathrm{aux}}$ collects rationale-conditioned utility prediction and lightweight source-specific regularizers such as future-action prediction. The compact objective and dataset-specific instantiations are given in Appendix~\ref{app:detailed_method}.

\subsection{Policy Learning under Mental-conditioned Reward}

To integrate the learned mental-model into a deployable agent, we use it as a training signal through the utility. Specifically, the policy must learn to map observable history directly to actions that the mental-model-aware utility would prefer. Through this training, the policy parameters internalize which observable contexts call for which mental-state-sensitive actions.

For the evaluation scalarizer $w_e\in\Delta^{D_e-1}$, the mental-model-aware utility induces a scalar reward
\begin{equation}
r_{\omega^{\star},\psi^{\star}}(h_t,a_t)
=
w_e^\top \bar{\mathbf{u}}_{\omega^{\star},\psi^{\star}}(h_t,a_t).
\label{eq:scalar_belief_reward_main}
\end{equation}

Policy learning then proceeds through candidate generation and mental-model-aware scoring. For a history $h_t$, the current policy proposes an action conditioned on $p_t=\phi_{\mathrm{pol}}(h_t)$; the scorer evaluates that action using Eq.~\eqref{eq:scalar_belief_reward_main}; and the policy update increases the likelihood of actions whose reward is high relative to other plausible choices in the same context. In implementation, we instantiate this update with mental-reward-guided GRPO, using $r_{\omega^\star,\psi^\star}$ to score sampled candidates and form group-relative advantages (Fig. 2). A compact objective for this update is
\begin{equation}
\phi^{\star}
\in
\arg\max_{\phi}
\mathbb{E}_{\tau\sim\pi_{\phi}}
\left[
\sum_t
r_{\omega^{\star},\psi^{\star}}(h_t,a_t)
-
\eta_{\mathrm{ref}}\,
\mathrm{KL}\!\left(
\pi_{\phi}(\cdot\mid p_t)\,\|\,\pi_{\mathrm{ref}}(\cdot\mid p_t)
\right)
\right].
\label{eq:policy_learning_main}
\end{equation}
The reference term keeps the update near the base policy while the mental-model-aware reward changes the relative preference among plausible actions. In open-ended episodes, the candidates are sampled responses. In structured ToM benchmarks, the same signal scores answer options or constrained outputs.

The key mechanism is gradient assignment. When an action receives high $r_{\omega^{\star},\psi^{\star}}$, the update increases $\log \pi_{\phi}(a_t\mid p_t)$ in contexts sharing the observable cues in $h_t$. Because that reward was computed by marginalizing over $q_{\omega^{\star}}(z_t\mid h_t)$, the gradient carries information about the hidden mental state. Over training, the policy absorbs regularities linking observable cues to partner beliefs, intents, and goals, and the actions the critic prefers under them.

After training, the deployment uses only the learned policy
\begin{equation}
a_t\sim\pi_{\phi^{\star}}(\cdot\mid p_t).
\label{eq:test_time_policy_clean}
\end{equation}
 Detailed architectural design, component choices, dataset-specific instantiations and implementation hyperparameters are deferred to Appendix~\ref{app:detailed_method}.

\section{Experimentation}
\label{sec:experimentation}
\noindent\textbf{Implementation details.}
All trained open-source agents use LoRA adapters on 7B-scale language or vision-language backbones. We use Qwen2.5-7B-Instruct as the primary language backbone, including LLaMA2-7B-Chat and Mistral-7B-Instruct-v0.3 as additional \textsc{Sotopia} comparisons, and use Qwen2.5-VL-Chat for multimodal experiments on MMRole. For \textsc{Sotopia}, we augment \textsc{Sotopia}-$\pi$\cite{WangEtAl2024SOTOPIAPi} episodes with ToM labels and utility annotations to train the mental-state/reward model; for BigToM and ToMi, we train on 5K synthetic BigToM-style samples and evaluate on the official \textsc{BigToM} and \textsc{ToMi} splits; for \textsc{MMRole}, we augment the released training set with mental labels. \textsc{Craigslist-bargain} and \textsc{FANToM} are used only as transfer evaluations. We first train the mental-state/reward model and then optimize the policy with supervised warmup when used, followed by mental-reward-guided GRPO using rewards from the frozen scorer. Detailed data construction, sample counts, latent dimensions, optimizer settings, decoding parameters, and hardware are reported in Appendix~\ref{app:implementation_details}; Appendix Table~\ref{tab:dataset-comparison} summarizes the augmented supervision, and Appendix~\ref{app:annotation-validation} reports automated filtering and blinded human verification.\\
\noindent\textbf{Baselines and evaluation.}
For \textsc{Sotopia}, we compare against zero-shot/base LLaMA2-7B, Mistral-7B, and Qwen2.5-7B agents, SFT and GRPO variants trained on our generated data, and closed-source GPT-4o-mini/Gemini baselines, evaluating with the official episode-level seven-dimension \textsc{Sotopia} scoring protocol on all, hard, and partner-averaged settings. Craigslist-Bargain uses the released negotiation protocol and reports deal, walkaway, and no-agreement rates. For BigToM and ToMi, we follow the official evaluation scripts and report the released TB/FB/TB$\land$FB, Average, Joint, and false-belief split metrics against GPT-style and classic ToMi baselines where available. For MMRole, we compare with the official closed-source/open-source VLM and MMRole-Agent baselines and report the benchmark's normalized role-playing dimensions and overall score.

\begin{table}[t]
  \centering
  \small

  \begin{minipage}[t]{0.56\linewidth}
    \vspace{0pt}
    \centering
    \caption{Aggregated performance on \textsc{Sotopia-all} episodes with a fixed GPT-4o-mini partner.}
    \label{tab:sotopia-all}
    \setlength{\tabcolsep}{3.5pt}
    \resizebox{\linewidth}{!}{
      \begin{tabular}{l c cc cc cc}
        \toprule
        \multirow{2}{*}{Dimension} & \multirow{2}{*}{Range}
        & \multicolumn{2}{c}{LLaMA}
        & \multicolumn{2}{c}{Mistral}
        & \multicolumn{2}{c}{Qwen} \\
        \cmidrule(lr){3-4} \cmidrule(lr){5-6} \cmidrule(lr){7-8}
        & & Base & Ours & Base & Ours & Base & Ours \\
        \midrule
        Social rules  & $[-10,0]$ & -0.200 & -0.167 & -0.122 & -0.189 & -0.144 & -0.122 \\
        Secret        & $[-10,0]$ & -0.511 & -0.256 & -0.156 & -0.111 & -0.144 & -0.133 \\
        Financial     & $[-5,5]$  & 0.522 & 0.622 & 0.489 & 0.833 & 0.333 & 0.844 \\
        Relationship  & $[-5,5]$  & 2.656 & 2.844 & 3.022 & 3.178 & 2.656 & 3.533 \\
        Knowledge     & $[0,10]$  & 5.333 & 5.933 & 5.722 & 6.456 & 5.144 & 6.611 \\
        Believability & $[0,10]$  & 8.678 & 8.722 & 8.787 & 8.856 & 8.722 & 8.933 \\
        Goal          & $[0,10]$  & 6.833 & 7.089 & 6.833 & 7.467 & 6.467 & 7.322 \\
        \midrule
        \textbf{Overall} & & 3.330 & \textbf{3.541} & 3.511 & \textbf{3.784} & 3.290 & \textbf{3.856} \\
        \bottomrule
      \end{tabular}
    }
  \end{minipage}\hfill
  \begin{minipage}[t]{0.44\linewidth}
    \vspace{0pt}
    \centering
    \begin{minipage}[c]{0.56\linewidth}
      \centering
      \includegraphics[width=\linewidth]{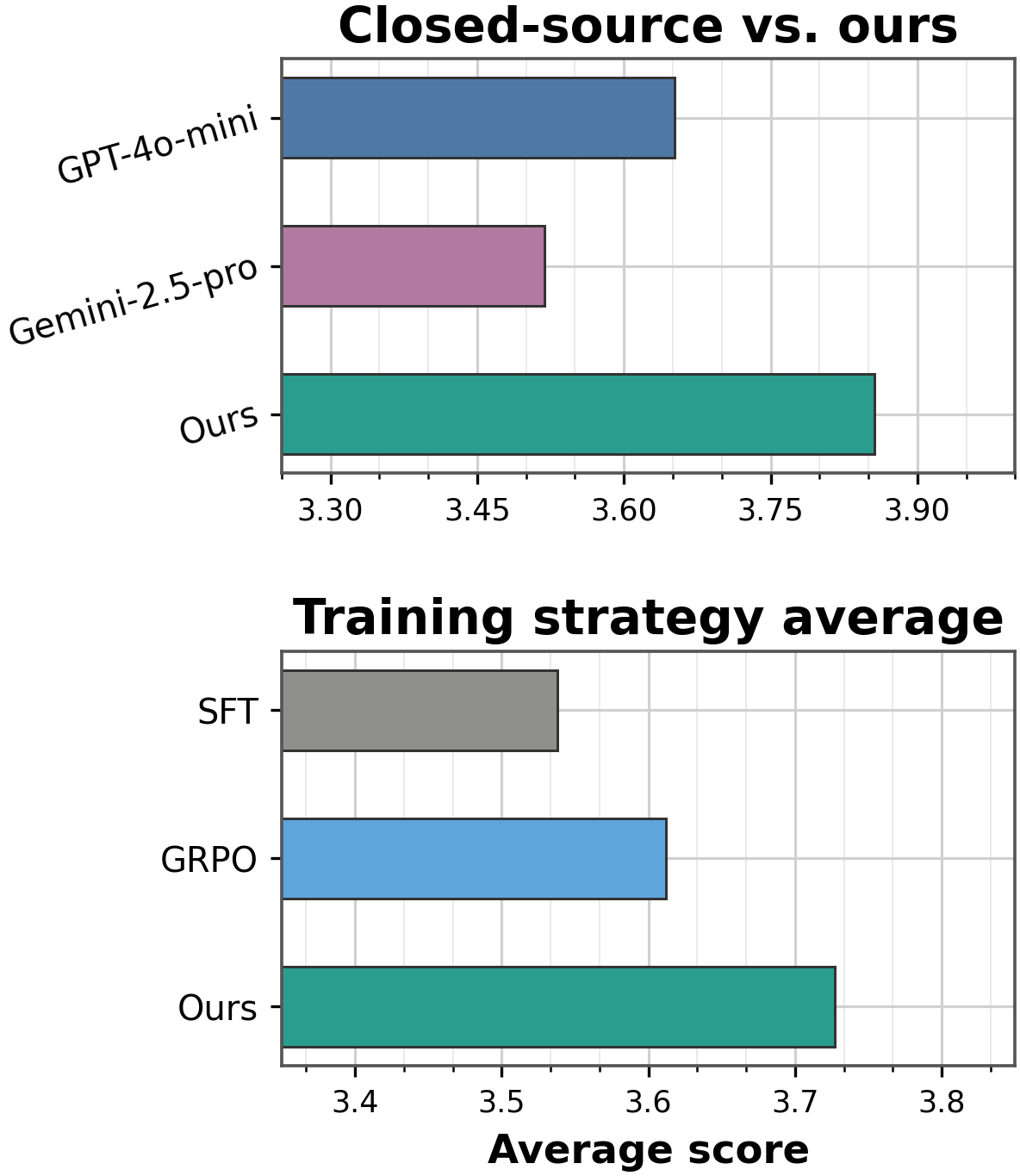}
    \end{minipage}\hfill
    \begin{minipage}[c]{0.41\linewidth}
      \captionof{figure}{\textsc{Sotopia} average-score comparisons. \textbf{Top:} closed-source models versus ours. \textbf{Bottom:} SFT, GRPO, and ours averaged over open-source models.}
      \label{fig:sotopia-bar-comparisons}
    \end{minipage}
  \end{minipage}
  \vspace{-0.9em}
\end{table}

\begin{table}[t]
  \centering
  \small
  \begin{minipage}[t]{0.31\linewidth}
    \vspace{0pt}
    \caption{Selected comparison on \textsc{Sotopia-hard} episodes with GPT-4o-mini partner.}
    \label{tab:hard-sotopia}
    \centering
    \setlength{\tabcolsep}{3pt}
    \resizebox{\linewidth}{!}{
      \begin{tabular}{llcccc}
        \toprule
        Model & Method & Goal & Bel. & Know. & Avg. \\
        \midrule
        \multirow{2}{*}{LLaMA} & Base & 5.000 & 8.571 & 4.071 & 5.880 \\
        & Ours & 5.571 & 8.500 & 4.429 & \textbf{6.166} \\
        \midrule
        \multirow{2}{*}{Mistral} & Base & 5.071 & 8.429 & 6.643 & 6.714 \\
        & Ours & 5.714 & 8.714 & 5.786 & \textbf{6.738} \\
        \midrule
        \multirow{2}{*}{Qwen} & Base & 5.643 & 8.643 & 4.857 & 6.381 \\
        & Ours & 5.643 & 9.000 & 5.714 & \textbf{6.785} \\
        \bottomrule
      \end{tabular}
    }
  \end{minipage}\hfill
  \begin{minipage}[t]{0.34\linewidth}
    \vspace{0pt}
    \caption{Partner-averaged (GPT-4o-mini and self) scores on selected \textsc{Sotopia} dimensions.}
    \label{tab:sotopia-partner-average}
    \centering
    \small
    \setlength{\tabcolsep}{3pt}
    \resizebox{0.96\linewidth}{!}{
      \begin{tabular}{llcccc}
        \toprule
        Model & Method & Goal & Bel. & Know. & Avg. \\
        \midrule
        \multirow{2}{*}{LLaMA} & Base & 6.333 & 8.561 & 4.872 & 6.588 \\
        & Ours & 7.045 & 8.739 & 5.950 & \textbf{7.245} \\
        \midrule
        \multirow{2}{*}{Mistral} & Base & 6.878 & 8.766 & 5.683 & 7.109 \\
        & Ours & 7.273 & 8.850 & 6.534 & \textbf{7.552} \\
        \midrule
        \multirow{2}{*}{Qwen} & Base & 6.773 & 8.811 & 5.678 & 7.087 \\
        & Ours & 7.317 & 8.945 & 6.639 & \textbf{7.633} \\
        \bottomrule
      \end{tabular}
    }
  \end{minipage}\hfill
  \begin{minipage}[t]{0.29\linewidth}
    \vspace{0pt}
    \caption{Craigslist-bargain generalization transfer statistics.}
    \label{tab:negotiation}
    \centering
    \small
    \renewcommand{\arraystretch}{1.0}
    \setlength{\tabcolsep}{2pt}
    \resizebox{\linewidth}{!}{
      \begin{tabular}{@{}llccc@{}}
        \toprule
        Model & Method & Deal & Walk. & No Agr.  \\
        \midrule
        \multicolumn{5}{@{}l}{\textit{Closed-Source}} \\
        GPT-4o-mini &  & 96.0\%   & 3.0\%   & 1.0\%    \\
        \midrule
        \multicolumn{5}{@{}l}{\textit{Open-Source (7B)}} \\
        \multirow{2}{*}{LLaMA2} & Base & 93.0\%   & 7.0\%   & 0.0\%   \\
        & Ours & 94.0\%   & 1.0\%   & 5.0\%    \\[1pt]
        \multirow{2}{*}{Qwen2.5} & Base & 91.0\%   & 7.0\%   & 2.0\%    \\
        & Ours & \textbf{94.0\%} & \textbf{5.0\%} & \textbf{1.0\%} \\
        \bottomrule
      \end{tabular}
    }
  \end{minipage}
  \vspace{-2em}
\end{table}

\subsection{Main Experimentation Results}

\noindent
\textbf{Results on \textsc{Sotopia}. }
Table~\ref{tab:sotopia-all} reports the main \textsc{Sotopia-all} evaluation results under a fixed GPT-4o-mini partner. Our method improves the overall score for all three open-source agents (LLaMA: 3.330$\to$3.541, Mistral: 3.511$\to$3.784, Qwen: 3.290$\to$3.856), indicating that the gains are not specific to one model family. The pronounced improvements on task-facing dimensions such as Knowledge and Goal, with high and stable Believability, suggests that the learned mental-model signal improves goal-directed reasoning without making the agents less plausible. Table~\ref{tab:hard-sotopia} shows the same pattern on \textsc{Sotopia-hard}, where average scores improve for all three LLM agents,
showing that the method continues to help when scenarios demand more difficult partner modeling.

Table~\ref{tab:sotopia-partner-average} evaluates generalization across partner types by averaging results over the GPT-4o-mini and self-partner settings. The improvement is consistent across all agents (LLaMA: $6.588 \to 7.245$, Mistral: $7.109 \to 7.552$, and Qwen: $7.087 \to 7.633$). These gains suggest that the method does not depend on a particular partner model and transfers across both closed-source and open-source counterparts.

Figure~\ref{fig:sotopia-bar-comparisons} summarizes the aggregate \textsc{Sotopia} trends. Our 7B Qwen2.5 agent with explicit mental-model training outperforms stronger closed-source baselines in the top panel, while the bottom panel shows consistent gains over SFT-only and GRPO-only variants (trained on our ToM-augmented data) averaged across LLaMA, Mistral, and Qwen. This suggests that explicit partner-state modeling contributes beyond standard supervised or reward-only training.

Finally, Table~\ref{tab:negotiation} evaluates cross-domain transfer. We evaluate the agents trained exclusively with \textsc{sotopia}-derived ToM data on a random subset of 100 \textsc{Craigslist Bargain} negotiation examples without any task-specific fine-tuning. The models retain high deal rates and reduce walkaways. Together, Tables~\ref{tab:sotopia-all}--\ref{tab:negotiation} and Figure~\ref{fig:sotopia-bar-comparisons} suggest that explicit partner-state modeling improves both in-domain social interaction and out-of-domain negotiation transfer.

\begin{table*}[t]
\centering
\begin{minipage}[t]{0.47\textwidth}
\vspace{0pt}
\centering
\caption{\textbf{\textsc{BigToM} results.} TB/FB/TB$\land$FB denote true-belief, false-belief, and paired-branch accuracy without/with the initial belief sentence. Bold marks the best open-source result.}
\label{tab:bigtom-main}
\tiny
\setlength{\tabcolsep}{2pt}
\renewcommand{\arraystretch}{0.92}
\resizebox{\linewidth}{!}{%
\begin{tabular}{llcccccc}
\toprule
& & \multicolumn{3}{c}{Without Initial Belief} & \multicolumn{3}{c}{With Initial Belief} \\
\cmidrule(lr){3-5} \cmidrule(lr){6-8}
{Task} & {Model} & TB & FB & TB$\land$FB & TB & FB & TB$\land$FB \\
\midrule
\multirow{7}{*}{Forward Belief}
& GPT-4      & 99.0 & 98.0 & 97.0 & 91.0 & 99.0 & 90.0 \\
& GPT-4 (CoT)  & 99.0 & 99.0 & 98.0 & 99.0 & 99.0 & 98.0 \\
\arrayrulecolor{gray!70}\cline{2-8}\arrayrulecolor{black}
& Base                & 83.0 & 94.0 & 77.0 & 49.5 & 98.5 & 48.0 \\
& SFT      & 95.5 & 88.5 & 84.5 & 94.0 & 96.5 & 92.0 \\
& Ours                & \textbf{97.0} & \textbf{98.0} & \textbf{96.0} & \textbf{97.5} & \textbf{99.5} & \textbf{97.5} \\
\midrule
\multirow{7}{*}{Forward Action}
& GPT-4      & 98.0 & 81.0 & 79.0 & 98.0 & 92.0 & 90.0 \\
& GPT-4 (CoT)  & 99.0 & 88.0 & 87.0 & 99.0 & 96.0 & 95.0 \\
\arrayrulecolor{gray!70}\cline{2-8}\arrayrulecolor{black}
& Base                & 99.5 & 64.0 & 63.5 & \textbf{99.5} & 68.0 & 67.5 \\
& SFT       & 93.0 & 82.5 & 76.5 & 92.0 & 80.0 & 73.0 \\
& Ours                & \textbf{99.5} & \textbf{79.0} & \textbf{78.5} & 99.0 & \textbf{81.0} & \textbf{80.0} \\
\midrule
\multirow{7}{*}{Backward Belief}
& GPT-4      & 86.0 & 53.0 & 40.0 & 62.0 & 77.0 & 40.0 \\
& GPT-4 (CoT)  & 84.0 & 54.0 & 38.0 & 76.0 & 63.0 & 40.0 \\
\arrayrulecolor{gray!70}\cline{2-8}\arrayrulecolor{black}
& Base                & 85.0 & 92.5 & 77.5 & 53.5 & 99.0 & 52.5 \\
& SFT       & 94.5 & 90.5 & 85.0 & 94.5 & 95.5 & 91.0 \\
& Ours                & \textbf{97.0} & \textbf{98.5} & \textbf{95.5} & \textbf{98.5} & \textbf{99.5} & \textbf{98.5} \\
\bottomrule
\end{tabular}
}
\end{minipage}\hfill
\begin{minipage}[t]{0.51\textwidth}
\vspace{5pt}
\centering
\caption{\textbf{\textsc{MMRole} results.} Official normalized role-play metrics are averaged into Overall. Blue rows mark ours, which improves over MMRole-Agent overall and on In/Out splits.}
\vspace{1em}
\label{tab:mmrole-benchmark}
\tiny
\setlength{\tabcolsep}{1.6pt}
\renewcommand{\arraystretch}{0.9}
\resizebox{\linewidth}{!}{%
\begin{tabular}{lcccccccc | c}
\toprule
MRPAs & IA & Flu & Coh & ITR & RA & PC & KC & TC & Overall \\
\midrule
GPT-4 Turbo            & 1.055 & 1.032 & 1.084 & 1.097 & 1.092 & 1.168 & 1.103 & 1.161 & 1.099 \\
Gemini Pro Vision      & 0.999 & 1.007 & 1.028 & 1.009 & 1.013 & 1.052 & 1.013 & 1.050 & 1.021 \\
Claude 3 Opus          & 1.127 & 1.070 & 1.149 & 1.167 & 1.146 & 1.219 & 1.168 & 1.213 & \textbf{1.157} \\
QWen-VL-Max            & 1.014 & 1.012 & 1.035 & 1.034 & 1.029 & 1.042 & 1.021 & 1.041 & 1.028 \\
\midrule
QWen-VL-Chat           & 0.844 & 0.954 & 0.879 & 0.850 & 0.829 & 0.778 & 0.827 & 0.785 & 0.843 \\
LLaVA-NeXT-Mistral-7B  & 0.948 & 0.986 & 0.964 & 0.938 & 0.933 & 0.924 & 0.940 & 0.921 & 0.944 \\
Yi-VL-6B               & 0.844 & 0.919 & 0.859 & 0.828 & 0.811 & 0.776 & 0.820 & 0.774 & 0.829 \\
MMRole-Agent           & 0.998 & 1.000 & 0.997 & 0.993 & 0.987 & 1.000 & 0.992 & 0.988 & 0.994 \\
\rowcolor{blue!8}
Ours         & 1.013 & 1.001 & 1.043 & 1.010 & 1.031 & 1.032 & 1.031 & 1.021 & \textbf{1.023} \\
\midrule
MMRole-Agent (In)  & 1.000 & 1.000 & 0.999 & 0.997 & 0.989 & 1.012 & 0.997 & 0.997 & 0.999 \\
MMRole-Agent (Out) & 0.992 & 0.999 & 0.993 & 0.979 & 0.981 & 0.963 & 0.977 & 0.962 & 0.981 \\
\rowcolor{blue!8}
Ours (In)         & 1.014 & 1.000 & 1.044 & 1.008 & 1.031 & 1.040 & 1.033 & 1.024 & \textbf{1.024} \\
\rowcolor{blue!8}
Ours (Out)        & 1.009 & 1.004 & 1.041 & 1.017 & 1.030 & 1.008 & 1.026 & 1.015 & \textbf{1.019} \\
\bottomrule
\end{tabular}%
}
\end{minipage}
\vspace{-0.8em}
\end{table*}

\begin{table*}[t]
\centering
\begin{minipage}[t]{0.63\textwidth}
\vspace{0pt}
\centering
\scriptsize
\setlength{\tabcolsep}{3pt}
\renewcommand{\arraystretch}{0.98}
\resizebox{\linewidth}{!}{%
\begin{tabular}{lcccccccc}
\toprule
Model & Average & Joint & \shortstack{w/o FB\\1st} & \shortstack{w/o FB\\2nd} & \shortstack{w/o FB\\Reality} & \shortstack{w/o FB\\Memory} & \shortstack{FB\\1st} & \shortstack{FB\\2nd} \\
\midrule
\underline{\textit{Classic TOMi baselines}}\\
MemNN  & 77.20 & 44.30 & 85.45 & 82.67 & 93.39 & 98.90 & 12.62 & 17.27 \\
RelNet  & 86.00 & 57.40 & 96.42 & 95.37 & 100.0 & 99.90 & 10.40 & 17.81 \\
EntNet  & \textbf{90.60} & 66.80 & 94.29 & 85.08 & 100.0 & 100.0 & 54.95 & 36.55 \\
\midrule
\underline{\textit{Closed source LLMs}}\\
GPT-3.5-Instruct  & 79.78  & 36.94 &94.42  &84.05  & 97.30 &94.79 & 34.26 & 7.82 \\
GPT-4o-mini  & 71.00 & 24.42 & 77.92 & 12.57 & 99.40 & 100.0 & 81.68 & 92.73 \\
\midrule
Base & 87.92 & 55.16 & 91.09 & 78.45 & 99.40 & 99.80 & 74.26 & 71.27 \\
\textbf{Ours} & \underline{89.16} & 60.46 & 93.16 & 91.09 & 94.79 & 98.20 & 66.34 & 62.55 \\
\bottomrule
\end{tabular}%
}
\end{minipage}\hfill
\begin{minipage}[t]{0.35\textwidth}
\vspace{-5pt}
\caption{\textbf{\textsc{ToMi} Generalization} from the \textsc{BigToM}-trained policy. Ours improves the aggregate Average and Joint scores over the base model and strengthens non-false-belief first-/second-order belief tracking, highlighting OOD generalization of our mental-model trained policy.}
\label{tab:tomi-neurips}
\end{minipage}
\vspace{-1em}
\end{table*}

\noindent
\textbf{Results on \textsc{BigToM} \& \textsc{ToMi}. }
Tables~\ref{tab:bigtom-main} \&~\ref{tab:tomi-neurips} evaluate transfer from open-ended dialogue to controlled ToM reasoning. \textsc{BigToM} (Table~\ref{tab:bigtom-main}) is especially diagnostic because TB$\land$FB only rewards a model when both branches of the same scenario are correct. The model must maintain two counterfactual mental states for the same agent. Base models often handle the easier true-belief branch but lose consistency once the false-belief branch or initial belief changes. SFT on our enriched data improves, but it still treats the output largely as a supervised answer pattern. In contrast, our mental-model variant is more consistent across forward belief, forward action, and backward belief because of its ability to encode higher order hidden states (further comparisons Appendix Table~\ref{tab:bigtom-main-supp}). 
\\
Table~\ref{tab:tomi-neurips} provides an OOD generalization check on \textsc{ToMi}, whose structure and metrics differ from the \textsc{BigToM} training format. We cross-evaluate our policy trained with \textsc{BigToM}-style data on \textsc{ToMi} and observe that it improves Average and Joint scores over the base model. The classic ToMi baselines remain strong because they are task-specific models fit to ToMi's synthetic state-tracking format and do not have broader multi-agent generality. Further, our model's high score on w/o FB second-order questions shows that the transferred latent is capturing nested belief structure when no false-belief conflict is introduced. Overall, the gains suggest that the model learns a reusable hidden-state interface for ToM.

\textbf{Results on \textsc{MMRole}. }
Table~\ref{tab:mmrole-benchmark} shows that our multimodal mental-model training improves over the released \textsc{MMRole}-Agent baseline on the overall score, both in-domain and out-of-domain. The gain is most visible on the Out split, where ours reaches $1.019$ against $0.981$ for \textsc{MMRole}-Agent (Out), indicating that the learned mental state does not merely fit seen character-image pairs but also transfers to held-out role contexts. The strongest improvements appear on Coherence, Response Accuracy, Personality Consistency, Knowledge Consistency, and Tone Consistency, which are precisely the dimensions where role-play depends on more than image recognition or fluent captioning. These metrics require inferring who the character is, what they know, how they should speak, and how the visual scene relates to the current dialogue.

\section{Analysis and Ablations}

\begin{figure}[t]
\vspace{-0.6em}
\centering
\begin{minipage}[t]{0.28\textwidth}
\centering
\includegraphics[width=\linewidth]{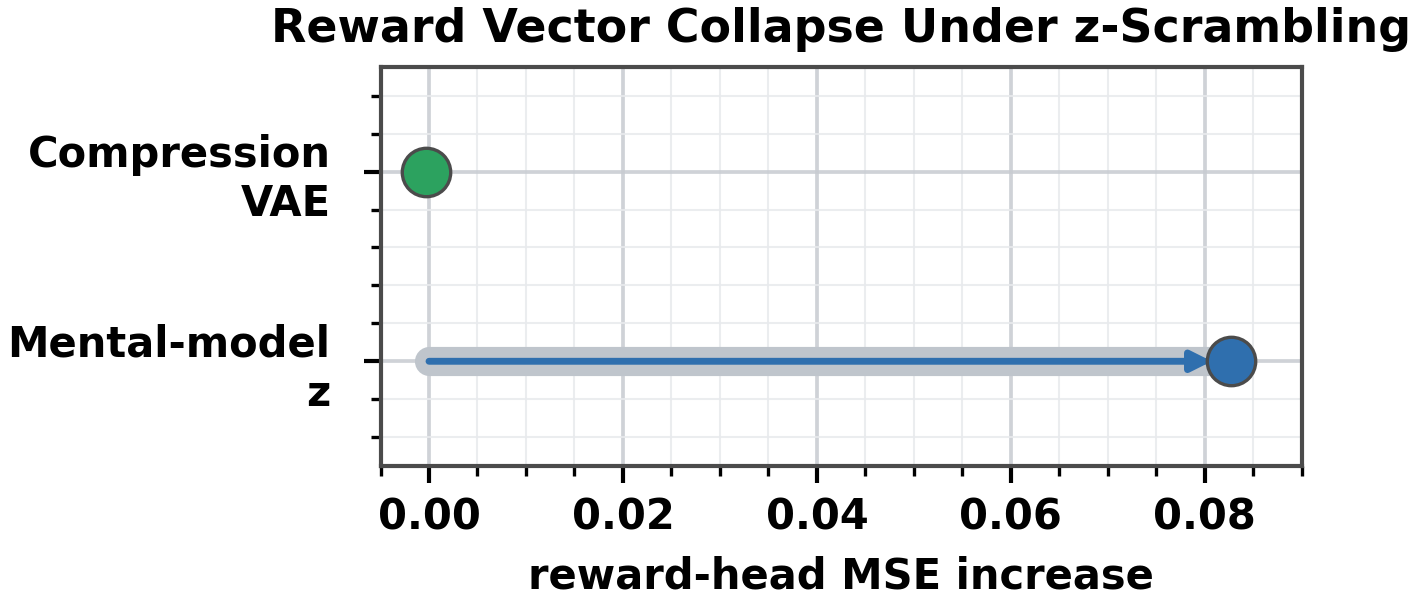}
\vspace{-0.5em}
\centerline{\scriptsize \textbf{a)}}
\end{minipage}\hfill
\begin{minipage}[t]{0.235\textwidth}
\centering
\includegraphics[width=\linewidth]{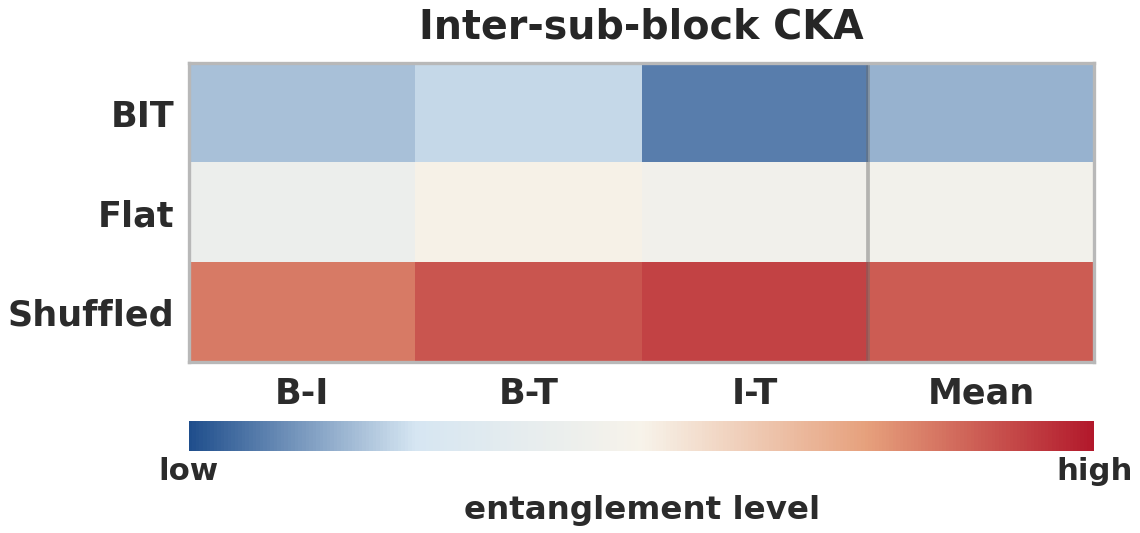}
\vspace{-0.5em}
\centerline{\scriptsize \textbf{b)}}
\end{minipage}\hfill
\begin{minipage}[t]{0.235\textwidth}
\centering
\includegraphics[width=\linewidth]{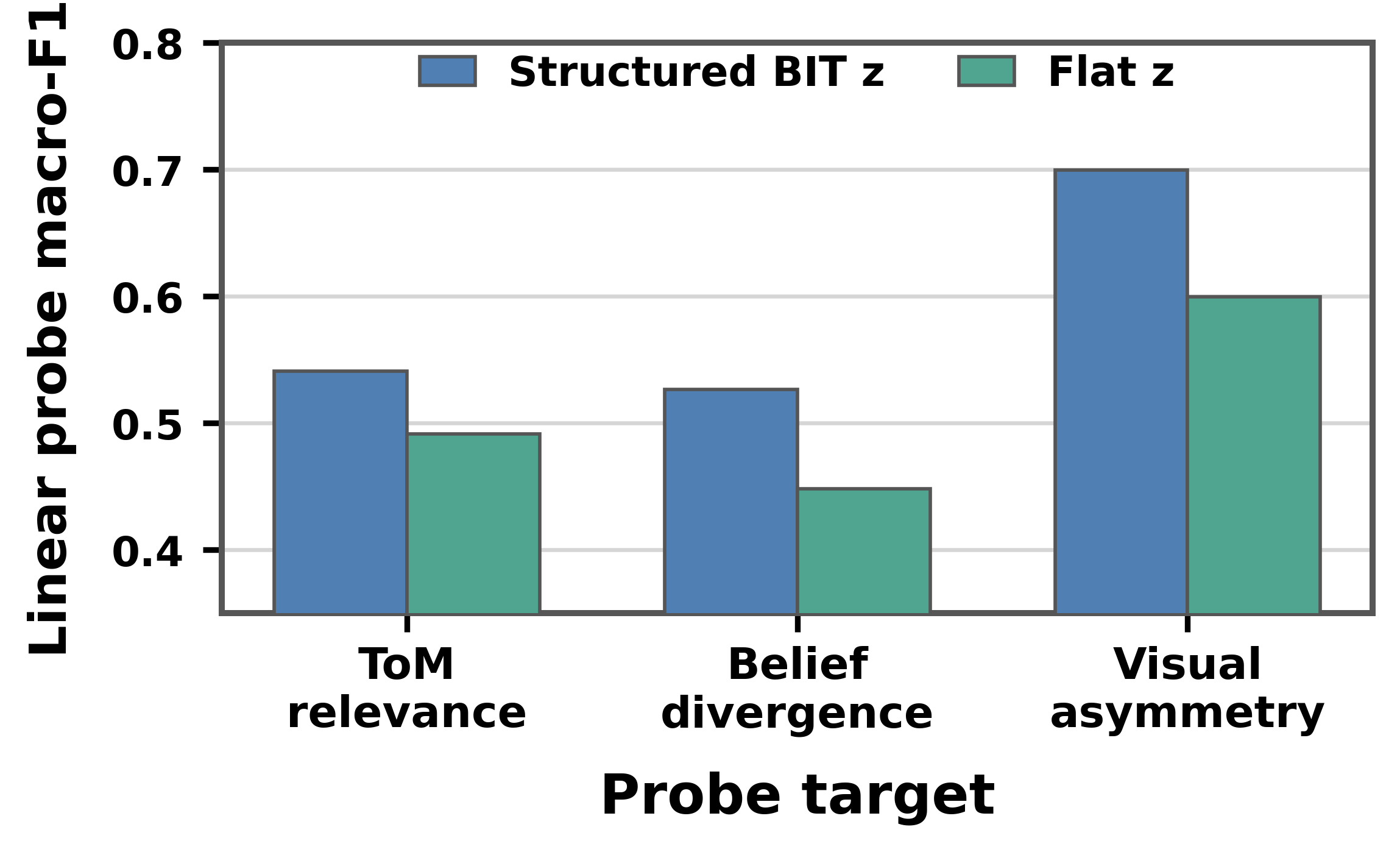}
\vspace{-0.5em}
\centerline{\scriptsize \textbf{c)}}
\end{minipage}\hfill
\begin{minipage}[t]{0.235\textwidth}
\centering
\includegraphics[width=\linewidth]{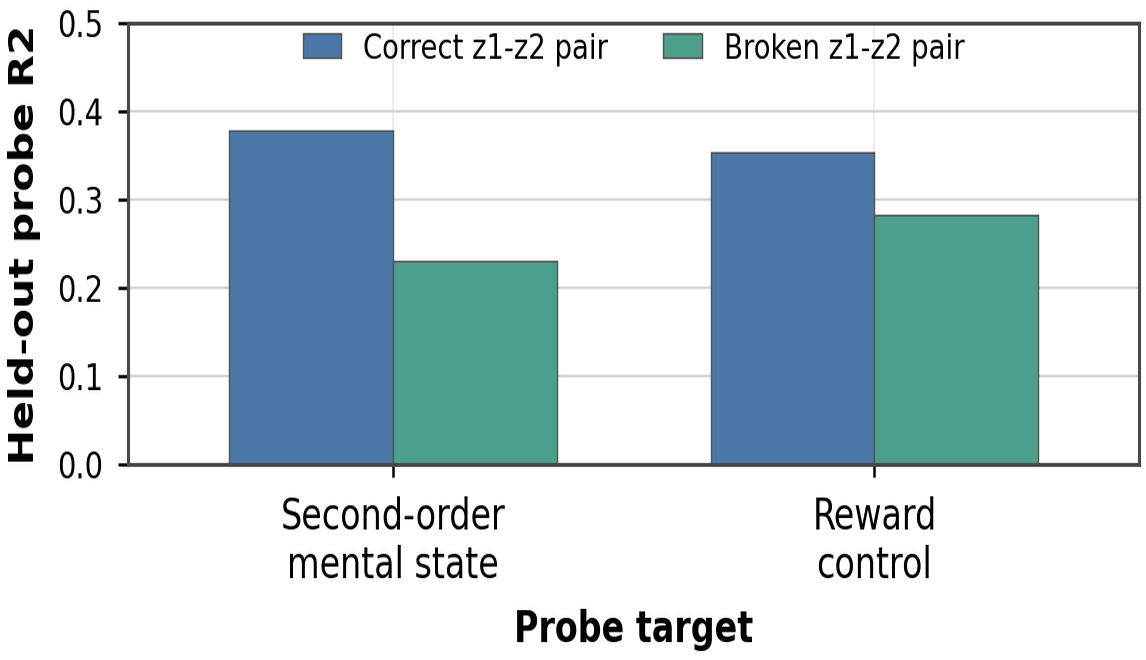}
\vspace{-0.5em}
\centerline{\scriptsize \textbf{d)}}
\end{minipage}
\caption{a) $z$-scrambling barely affects the compression VAE but sharply increases BIT reward-vector MSE, showing stronger reward-causal structure. b) Pairwise CKA shows complementary belief, intent, and thought slices. c) Frozen-latent probes recover ToM-relevant \textsc{MMRole} factors more separably from BIT than from a flat latent. d) Shuffling $z_2$ weakens second-order probes, supporting the recursive $z_1 \rightarrow z_2$ structure.}
\label{fig:reward-vector-collapse}
\label{fig:flat-vs-bit-latent-probe}
\label{fig:latent-pairing-probe}
\vspace{0em}
\end{figure}

\noindent
\textbf{Is mental-model simply a compression code? }
We train a flat compression VAE on the same ToM-augmented \textsc{Sotopia}-$\pi$ contexts and responses, and compare it our recursive mental-model. We fit lightweight probes from $z$ to the reward head. We then scramble $z$ across held-out examples while keeping the reward head fixed. Figure~\ref{fig:reward-vector-collapse}a shows that scrambling the flat compression VAE latent has only a weak effect on reward-vector MSE, while scrambling mental-model $z$ sharply disrupts joint reward prediction. Thus, the flat latent can store reconstructive mental-summary information, but the mental-model $z$ latent is more tightly coupled to reward-relevant decision structure.

\textbf{Significance of BIT structure. }
Figure~\ref{fig:reward-vector-collapse}b probes the internal structure of the belief, intent, and thought (BIT) slices using pairwise CKA on \textsc{Sotopia} data. The BIT variant produces subspaces that are separated yet related, whereas the model trained with flat mental-summary including shuffled variant shows significant inter-slice similarity. Figure~\ref{fig:reward-vector-collapse}c further shows that linear probes recover ToM-relevant \textsc{MMRole} factors more accurately from BIT than from a flat latent of the same combined dimensionality. Together, these diagnostics support the intended interpretation that BIT slices are complementary components rather than redundant copies.

\textbf{Recursive $z_1 \rightarrow z_2$ structure. }
Figure~\ref{fig:reward-vector-collapse}d tests whether the relation between first- and second-order latents carries intrinsic mental-state information, rather than simply adding another independent code. We compare probes on the true example-wise pair $(z_1,z_2)$ against samples where $z_2$ is shuffled while preserving its marginal distribution. This keeps the amount of latent capacity and the statistics of each block largely unchanged, but breaks the recursive alignment between a partner state and the higher-order state conditioned on it. The shuffled-pair control loses substantially more recoverable second-order mental-state information than reward information, suggesting that the $z_1 \rightarrow z_2$ link specifically carries nested ToM structure rather than only generic reward-relevant compression.

\noindent
\textbf{Sensitivity to mental-state supervision. }
We study whether the gains also depend on the available mental-state supervision. We train Qwen2.5-7B with nested $25\%$ and $50\%$ subsets of the \textsc{Sotopia} annotations (the $25\%$ subset is contained in the $50\%$ subset) while keeping the mental-model architecture fixed; $100\%$ uses all annotations, whereas $0\%$ is the mental-free reward baseline because no first- or second-order mental targets are supplied. Every condition uses the same evaluation seed, GPT-4o-mini partner, GPT-4o judge, and official \textsc{Sotopia} scoring protocol. Table~\ref{tab:supervision-fraction-main} shows that only $25\%$ supervision already improves the overall score by $0.350$ points over the baseline, and performance increases with annotation coverage to $3.856$. Thus, the method indeed benefits from limited supervision rather than requiring exhaustive annotation.

\noindent
\textbf{Posterior-family ablation. }
We study whether our proposed diagonal posterior loses important within-level correlations among belief, intent, and thought coordinates. We retrain the Stage-1 mental-state/reward model on \textsc{BigToM}, holding all settings fixed and replacing only the diagonal Gaussian with a structured full-covariance posterior, $\Sigma=\mathrm{diag}(\sigma^2)+UU^\top$. On the held-out scenario-level split ($n{=}5{,}976$), pairwise reward accuracy and margin measure preferred-versus-negative ranking, while the belief probe measures mental-state recoverability from the latent. Table~\ref{tab:posterior-ablation-main} shows nearly identical reward ranking, margin, and belief recovery; the $0.2$-point accuracy difference is within the $\pm0.65$-point sampling uncertainty at this $n$. The more costly covariance therefore provides no measurable benefit. Cross-coordinate dependence remains available through the shared nonlinear history encoder, the conditional $q(z_2\mid h,z_1)$, and the joint prediction heads; the diagonal assumption restricts only residual posterior noise.

\begin{table*}[t]
\centering
\small
\begin{minipage}[t]{0.36\textwidth}
\centering
\caption{Sensitivity to mental-state supervision on \textsc{Sotopia} with Qwen2.5-7B. Higher is better.}
\label{tab:supervision-fraction-main}
\setlength{\tabcolsep}{5pt}
\begin{tabular}{lc}
\toprule
Mental supervision & Overall score \\
\midrule
$0\%$ (baseline) & $3.300$ \\
$25\%$ & $3.650$ \\
$50\%$ & $3.750$ \\
$100\%$ & $\mathbf{3.856}$ \\
\bottomrule
\end{tabular}
\end{minipage}\hfill
\begin{minipage}[t]{0.61\textwidth}
\centering
\caption{Posterior-family ablation on a held-out, scenario-level \textsc{BigToM} split ($n{=}5{,}976$). Accuracy and probe scores are percentages.}
\label{tab:posterior-ablation-main}
\setlength{\tabcolsep}{4pt}
\resizebox{\linewidth}{!}{%
\begin{tabular}{lccc}
\toprule
Mental-model posterior & Reward acc. & Margin & Belief probe \\
\midrule
Diagonal Gaussian (ours) & $96.6$ & $1.05$ & $99.8$ \\
Full covariance, low-rank ($\mathrm{diag}+UU^\top$) & $96.8$ & $1.05$ & $99.9$ \\
\bottomrule
\end{tabular}%
}
\end{minipage}
\vspace{-0.5em}
\end{table*}

\begin{figure}[t]
\vspace{-0.6em}
\centering
\begin{minipage}[t]{0.42\textwidth}
  \centering
  \vspace{0pt}
  \includegraphics[width=\linewidth]{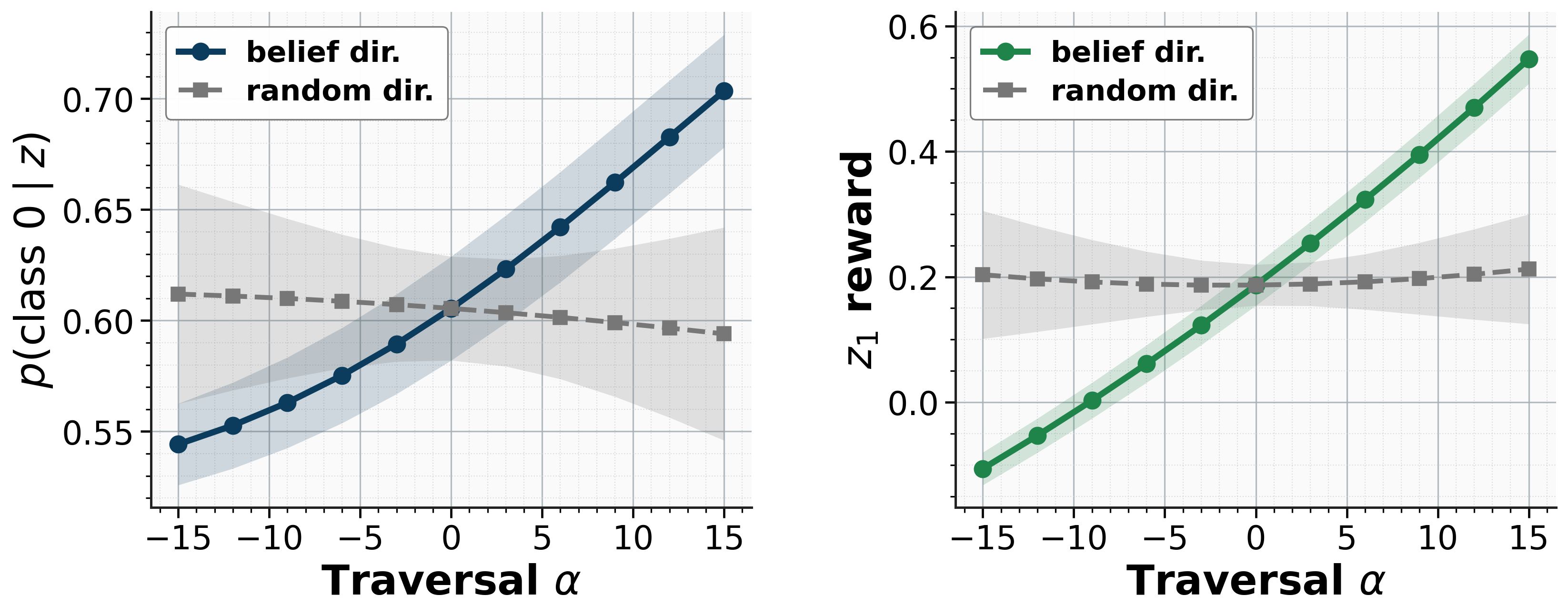}\\[-0.2em]
  \textbf{a)}
\end{minipage}\hfill
\begin{minipage}[t]{0.57\textwidth}
  \centering
  \vspace{0pt}
  \includegraphics[width=\linewidth]{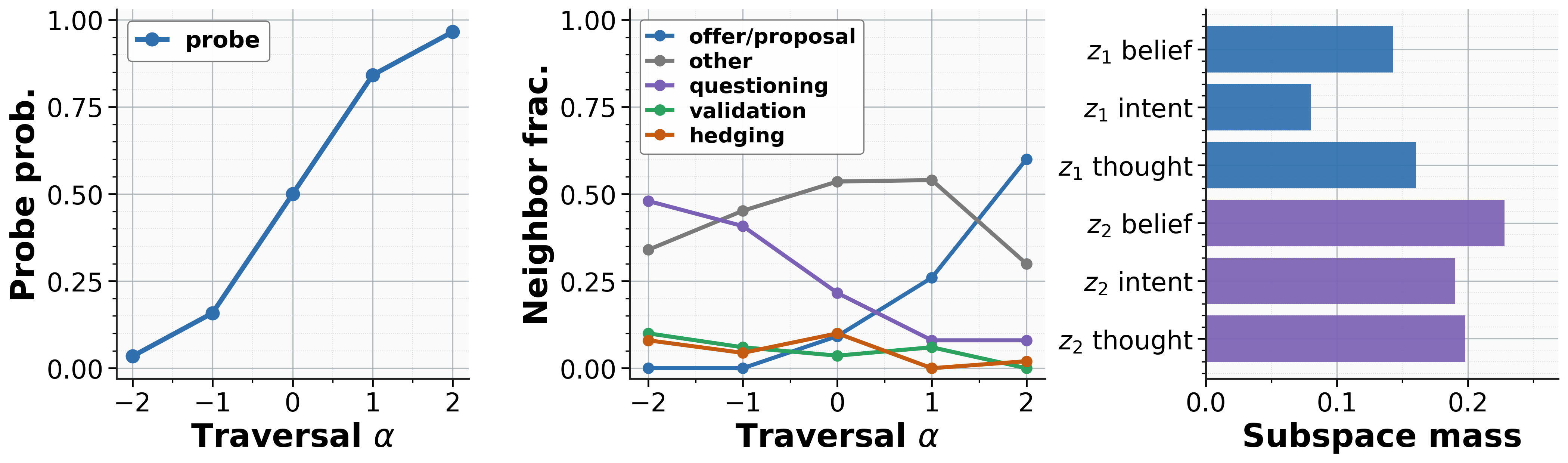}\\[-0.2em]
  \textbf{b)}
\end{minipage}
\vspace{-0.5em}
\caption{Latent traversal analyses. a) Belief-axis traversal shifts both the branch classifier and frozen reward head beyond random controls. b) Offer/proposal traversal increases strategy signal and local label enrichment, with stronger second-order contribution.}
\label{fig:latent-traversals}
\label{fig:belief-traversal}
\label{fig:offer-proposal-traversal}
\vspace{-1.0em}
\end{figure}

\noindent
\textbf{Functional Latent Traversals. }
We test whether recovered latent directions correspond to meaningful mental and strategic variables, rather than only post-hoc probe artifacts. For the belief analysis, we linearly probe a held-out branch label indicating whether an agent is aware or not-aware of an event, then perturb near-boundary anchors along the resulting direction $\hat{u}_{\text{belief}}$ and compare against magnitude-matched random directions. Figure~\ref{fig:latent-traversals}a shows that this traversal monotonically shifts both the branch classifier and $z_1$-only reward head, even though the reward head was not trained on the branch label. 
For the strategy analysis (Figure~\ref{fig:latent-traversals}b), we fit a linear probe for the annotated \textsc{Sotopia} strategy label \emph{offer/proposal} and move near-boundary latent points along this direction with traversal coefficient $\alpha$. This is not an edit to one generated response; it tests how the local latent neighborhood changes as $\alpha$ moves toward the strategy concept. In Figure~\ref{fig:latent-traversals}b, left panel shows the offer/proposal probe score increasing with $\alpha$, the middle panel shows nearest-neighbor labels shifting from questioning/other toward offer/proposal, and the right panel decomposes the direction into squared-weight mass across BIT blocks. The larger mass in second-order ToM blocks suggests that proposal-like strategies rely more on modeling how the partner may respond.

\section{Conclusion}
\label{sec:conclusion}
We introduced a mental-model-enabled agent framework for multi-agent systems, where an agent learns an amortized recursive Theory-of-Mind latent state and couples it to a belief-conditioned reward model. The resulting reward signal is used to train policies that act independently at test time while retaining the benefits of explicit partner-state modeling. Across \textsc{Sotopia}, negotiation transfer, \textsc{BigToM}/\textsc{ToMi}/\textsc{FANToM}, and \textsc{MMRole}, the same framework improves open-source agents on social-interaction and Theory-of-Mind benchmarks. Our results also show that augmenting existing datasets with mental-state annotations, utility signals, rationales, and contrastive negatives provides useful supervision for learning reusable partner representations.

\noindent\textbf{Limitations and future work.}
 Our training environments are primarily dyadic; the zero-shot \textsc{FANToM} transfer in Appendix~\ref{app:fantom-transfer} provides initial multi-party evidence, while end-to-end multi-party training with participant-specific latent updates remains future work. The same formulation can also be extended beyond language and image-grounded role play to embodied and robotic settings, where partner beliefs, goals, visibility, and physical constraints evolve over longer horizons. Finally, scaling to larger policies, more diverse partner models, richer online belief updates, and human-in-the-loop evaluation would help test how far explicit mental-state structure continues to improve robust multi-agent behavior.
\bibliographystyle{plain}
\bibliography{refs}

\clearpage
\appendix
\begin{center}
{\Large\bfseries Mental-Models for Multi-Agent Systems}
\end{center}

\section{Appendix}

The appendix is organized as below:
\begin{itemize}
    \item \textbf{A.1 Methodology, architecture, and training details} (Sec.~\ref{app:detailed_method})
    \item \textbf{A.2 Dataset-specific instantiations} (Sec.~\ref{app:dataset_instantiations})
    \item \textbf{A.3 Additional results and ablations} (Sec.~\ref{app:additional_results_ablations})
    \item \textbf{A.4 Impact Statement} (Sec.~\ref{app:impact-statement})
\end{itemize}

\subsection{Methodology, Architecture, and Training Details}
\label{app:detailed_method}

\subsubsection{Dataset-Specific Methodology and Notation}
\label{app:method_notation}

The main paper presents the benchmark-agnostic version of our method. Here we give the concrete instantiations used for each dataset. Across all benchmarks, an example is represented as a tuple
\begin{equation}
(c_i, a_i^{+}, a_i^{-}, \mathbf{u}_i, \boldsymbol{\gamma}_i, m_i^{(1)}, m_i^{(2)}, \xi_i),
\end{equation}
where $c_i$ is the observable context, $a_i^{+}$ is the observed or preferred response, $a_i^{-}$ is an optional hard negative, $\mathbf{u}_i$ is a dataset-specific utility vector, $\boldsymbol{\gamma}_i$ is an optional dimension mask indicating which utility dimensions a contrastive negative is intended to violate, $m_i^{(1)}$ and $m_i^{(2)}$ are first- and second-order mental-state targets when available, and $\xi_i$ is optional rationale or reward-explanation text. Missing fields are handled by masking the corresponding auxiliary losses, which lets the same objective support open-ended dialogue, structured belief QA, and multimodal role-play.

\subsubsection{Architecture Overview}
\label{app:architecture_overview}

Stage 1 uses a single causal language-model or vision-language backbone with low-rank adapters and task-specific heads. The backbone encodes the task context $c_i$ and, when scoring a candidate action, the isolated action $a_i$ in separate forward passes, producing pooled context and action states that feed the latent posterior and reward heads. In our implementation, both recursive latent levels use $d_z=128$ dimensions and are partitioned into belief, intent, and thought subspaces. These subspaces are not required by the general objective, but they provide a stable interface for attaching first-order and second-order mental-state decoders.

\subsubsection{Implementation Details}
\label{app:implementation_details}

All open-source models are trained with LoRA adapters on 7B-scale backbones. For language-only experiments, Qwen2.5-7B-Instruct is the main policy and reward backbone; LLaMA2-7B-Chat and Mistral-7B-Instruct-v0.3 are included for additional \textsc{Sotopia} comparisons. For multimodal experiments, we use Qwen2.5-VL-Chat. The training sources are: \textsc{Sotopia}-$\pi$ episodes augmented with ToM labels and utility annotations, 5K synthetic \textsc{BigToM}-style scenarios with first- and second-order mental labels, and \textsc{MMRole} inter-role training turns augmented with visual ToM labels. \textsc{TOMi}, official \textsc{BigToM}, official \textsc{Sotopia}, \textsc{Craigslist-bargain}, and \textsc{FANToM} are used only for evaluation or transfer checks. \textsc{Sotopia} training episodes do not overlap with the original \textsc{Sotopia} evaluation episodes.

Across benchmarks, we optionally use supervised warmup or mental-prefix SFT before GRPO. Unless otherwise noted, the latent mental state uses $d_z=128$ with belief, intent, and thought subspaces. Training uses bf16 mixed precision, AdamW, warmup scheduling, KL annealing, and delayed second-order latent updates. Typical hyperparameter ranges are LoRA rank $8$--$32$, LoRA $\alpha=16$--$64$, dropout $0.05$, reward-model learning rate $2{\times}10^{-5}$, GRPO learning rate $1{\times}10^{-6}$--$3{\times}10^{-6}$, group size $8$, clipping coefficient $0.1$--$0.2$, KL coefficient $0.06$--$0.1$, temperature $0.6$--$0.8$, and top-$p$ $0.9$--$0.95$. Experiments were run on NVIDIA RTX A6000 48GB GPUs.

\subsubsection{Shared Mental--Reward Objective}
\label{app:shared_objective}

Stage 1 learns a \emph{coupled} latent model of mental state and utility. The objective is not to infer hidden state in isolation, nor to fit a reward predictor independently of social reasoning, but to learn a shared latent representation that is simultaneously constrained by \emph{mental-state supervision} and \emph{action-utility supervision}. The datasets differ in what counts as context, response, utility, and mental-state annotation; the loss template below is shared.

A shared language-model or vision-language backbone encodes $c_t$ into a context representation $e(c_t)$ and any candidate action $a$ into an action representation $e(a)$. Dataset-specific subsections below define the concrete form of $c_t$, $\mathbf{u}_t$, mental-state targets, and negatives.

\paragraph{Recursive amortized mental-state inference.}
We parameterize the amortized belief with a recursive two-level latent hierarchy,
\begin{align}
q_{\omega}(z_t^{(1)}, z_t^{(2)} \mid c_t)
&=
q_{\omega_1}(z_t^{(1)} \mid c_t)\,
q_{\omega_2}(z_t^{(2)} \mid c_t, z_t^{(1)}), \\
q_{\omega_1}(z_t^{(1)} \mid c_t)
&=
\mathcal{N}\!\big(\mu_1(e(c_t)), \operatorname{diag}(\sigma_1^2(e(c_t)))\big), \\
q_{\omega_2}(z_t^{(2)} \mid c_t, z_t^{(1)})
&=
\mathcal{N}\!\big(\mu_2([e(c_t); z_t^{(1)}]), \operatorname{diag}(\sigma_2^2([e(c_t); z_t^{(1)}]))\big).
\end{align}
The first latent $z_t^{(1)}$ captures first-order inferences about the counterpart, while the second latent $z_t^{(2)}$ captures higher-order inferences conditioned on $z_t^{(1)}$. In our implementation, each latent may be further structured into semantically designated subspaces corresponding to factors such as belief, intent, and thought, but the methodology only requires that the posterior be recursively factorized.

\paragraph{Belief-aware utility prediction.}
Given a candidate action $a$, the main utility head predicts a vector-valued outcome
\begin{equation}
\hat{\mathbf{u}}_t^{\mathrm{joint}}(a)
=
g_{\psi}^{\mathrm{joint}}([z_t^{(1)}; z_t^{(2)}; e(a)])
\in \mathbb{R}^{D}.
\label{eq:joint_utility_expanded}
\end{equation}
This head is action-dependent and captures how the predicted utility changes with the candidate action under the inferred mental state.

\paragraph{Mental-state and rationale supervision.}
When textual mental-state supervision is available, we decode first- and second-order descriptions directly from the latent states:
\begin{equation}
p_{\eta_1}(m_t^{(1)} \mid z_t^{(1)}),
\qquad
p_{\eta_2}(m_t^{(2)} \mid z_t^{(2)}).
\label{eq:mental_decoders_expanded}
\end{equation}
In the implementation, the decoders project latent subspaces into learned prefix tokens and decode the mental-state text through cross-attention. These terms encourage $z_t^{(1)}$ and $z_t^{(2)}$ to encode interpretable social structure rather than serving as arbitrary latent codes.

When rationale or reward-explanation text $\xi_t$ is available, a small explanation-conditioned head combines $[z_t^{(1)};z_t^{(2)}]$ with its encoded tokens $H(\xi_t)$ and predicts the same utility target. This directly tests whether the textual rationale is consistent with the outcome rather than treating it as free-form generation alone.

\paragraph{Training objective.}
We display only the core terms supported by the leave-one-out analysis:
\begin{equation}
\mathcal{L}_{\mathrm{MR}}
=
\mathcal{L}_{\mathrm{vec}}
+
\lambda_p\mathcal{L}_{\mathrm{pref}}
+
\lambda_m\mathcal{L}_{\mathrm{mental}}
+
\lambda_e \mathcal{L}_{\mathrm{expl}}
+
\mathcal{L}_{\mathrm{KL}}.
\label{eq:stage1_objective_expanded}
\end{equation}
Here $\mathcal{L}_{\mathrm{vec}}$ is Huber regression from the action-conditioned utility head to the dataset target, and $\mathcal{L}_{\mathrm{mental}}$ reconstructs the two supervised mental states:
\begin{align}
\mathcal{L}_{\mathrm{vec}}
&=\mathbb{E}\!\left[
\ell_{\mathrm{Huber}}\!\left(
\hat{\mathbf{u}}_t^{\mathrm{joint}}(a_t^{+}),\mathbf{u}_t
\right)\right], \\
\mathcal{L}_{\mathrm{mental}}
&=-\mathbb{E}\!\left[
\log p_{\eta_1}(m_t^{(1)}\mid z_t^{(1)})
+\log p_{\eta_2}(m_t^{(2)}\mid z_t^{(2)})
\right].
\end{align}
$\mathcal{L}_{\mathrm{expl}}$ is the corresponding Huber loss from the explanation-conditioned head to $\mathbf{u}_t$. Each supervised term is masked when its target is unavailable.

The preference term ranks the observed or preferred action above a hard negative only on dimensions that the negative was constructed to violate. With $\gamma_{t,d}\in\{0,1\}$ denoting that mask,
\begin{equation}
\mathcal{L}_{\mathrm{pref}}
=
\mathbb{E}
\left[
\mathbf{1}[a_t^{-}\ \text{exists}]
\cdot
\frac{1}{\sum_{d=1}^{D}\gamma_{t,d}+\varepsilon}
\sum_{d=1}^{D}
\gamma_{t,d}
\operatorname{softplus}
\Big(
\hat u_{t,d}^{\mathrm{joint}}(a_t^{-})
-
\hat u_{t,d}^{\mathrm{joint}}(a_t^{+})
\Big)
\right].
\label{eq:pref_loss_expanded}
\end{equation}
When a dataset provides a scalar preference rather than dimension-specific negative metadata, the mask is all ones. For generated hard negatives, the mask prevents fluent or otherwise unviolated dimensions from being penalized solely because the candidate is negative.

The KL term regularizes the first-order posterior toward a unit Gaussian and the second-order posterior toward a learned prior conditioned on $z_t^{(1)}$:
\begin{equation}
\mathcal{L}_{\mathrm{KL}}
=
\beta_1\operatorname{KL}\!\left(q_{\omega_1}(z_t^{(1)} \mid c_t)\,\|\,\mathcal{N}(0,I)\right)
+
\beta_2\operatorname{KL}\!\left(q_{\omega_2}(z_t^{(2)} \mid c_t,z_t^{(1)})\,\|\,p_{\chi}(z_t^{(2)}\mid z_t^{(1)})\right).
\end{equation}
where $p_{\chi}(z_t^{(2)}\mid z_t^{(1)})$ is a diagonal Gaussian conditional prior. This keeps $z_t^{(2)}$ recursively tied to $z_t^{(1)}$ rather than imposing unconditional independence between the two latent levels.

This compact presentation follows our matched-checkpoint \textsc{Sotopia} ablation: removing mental-state reconstruction or rationale conditioning reduces reward-regression correlation by $29\%$ and $26\%$, respectively. Removing hard-negative training has a smaller effect on regression correlation ($7\%$) but lowers hard-negative preference accuracy from $100\%$ to $81\%$, confirming that it supplies a distinct discriminative signal.

\paragraph{Optional regularizers.}
We also optionally add latent-only utility predictions $\hat{\mathbf{u}}_t^{(1)}=g_{\psi}^{(1)}(z_t^{(1)})$ and $\hat{\mathbf{u}}_t^{(12)}=g_{\psi}^{(12)}([z_t^{(1)};z_t^{(2)}])$, first-token future-action prediction, and source-specific auxiliary classifiers or probe regressors. These terms provide only small incremental validation gains relative to the core objective above and primarily serve as anti-bypass regularizers or diagnostics; we therefore mention them only in text rather than expanding additional losses. They are activated only when the corresponding annotation is available.

Overall, this stage produces a latent state that is simultaneously
\emph{(i)} amortized from history,
\emph{(ii)} semantically grounded in explicit mental-state supervision when available,
and
\emph{(iii)} directly tied to downstream decision quality through coupled utility prediction.
Because missing annotations simply mask the corresponding auxiliary terms, the same training objective can be used across heterogeneous datasets with different supervision granularity.

\paragraph{Stabilization.}
We use three implementation stabilizers when training the recursive latent model: optional pre-warm training for the first-order latent encoder/decoder on external mental-reasoning data, KL annealing with delayed activation for the second-order latent, and an initial stop-gradient from $z^{(1)}$ into $z^{(2)}$ so that first-order reasoning stabilizes before higher-order coupling is introduced.

\subsubsection{Policy Optimization Details}
\label{app:policy_optimization_details}
\label{app:shared_policy_optimization}

After learning the mental-state and reward model, we freeze it and train a policy $\pi_{\phi}(a \mid p)$, where $p=\phi_{\mathrm{pol}}(h)$ is the policy input. We optionally initialize the policy with supervised fine-tuning on observed actions,
\begin{equation}
\mathcal{L}_{\mathrm{SFT}}
=
-\sum_i \log \pi_{\phi}(a_i^{+} \mid p_i).
\label{eq:sft_clean}
\end{equation}
Policy training distinguishes the generation prompt $p_i$ from the reward context $c_i$: the former is optimized for action generation, while the latter is optimized for latent inference and reward estimation. This split prevents reward estimation from becoming tied to prompt-engineering artifacts.

\paragraph{Scalar reward induced by the frozen model.}
Because the learned utility is vector-valued, we use a scalarization operator
\begin{equation}
\mathrm{Agg}_{w}(\mathbf{u}) = w^\top \mathbf{u},
\qquad
w \in \Delta^{D-1},
\label{eq:scalarization_clean}
\end{equation}
to map $\mathbf{u} \in \mathbb{R}^{D}$ to a scalar control signal. For a sampled action $a$, we define the response-dependent score
\begin{equation}
r(h,a)
=
\mathrm{Agg}_{w}\!\big(\hat{\mathbf{u}}^{\mathrm{joint}}(h,a)\big),
\end{equation}
and the context-only latent score
\begin{equation}
b(h)
=
\frac{1}{2}
\left[
\mathrm{Agg}_{w}\!\big(\hat{\mathbf{u}}^{(1)}(h)\big)
+
\mathrm{Agg}_{w}\!\big(\hat{\mathbf{u}}^{(12)}(h)\big)
\right].
\label{eq:prior_bonus_clean}
\end{equation}
The response-dependent term ranks candidate actions. The context-only score $b(h)$ is useful for diagnostics and optional prompt filtering, but it is not added to candidate-level GRPO advantages because it is constant across the $K$ samples for a fixed prompt and would cancel under group normalization.
During policy optimization, the frozen mental-state model is evaluated deterministically using posterior means for $z^{(1)}$ and $z^{(2)}$.

\paragraph{Group-relative policy optimization.}
For each policy input $p_i$, we sample $K$ candidate actions
\begin{equation}
a_{i,1}, \dots, a_{i,K} \sim \pi_{\phi}(\cdot \mid p_i).
\end{equation}
Their response-level advantages are normalized within the group:
\begin{equation}
A^{\mathrm{resp}}_{i,k}
=
\frac{r_{i,k} - \mu_i}{\sigma_i + \varepsilon},
\qquad
\mu_i = \frac{1}{K}\sum_{k=1}^{K} r_{i,k},
\qquad
\sigma_i^2 = \frac{1}{K}\sum_{k=1}^{K}(r_{i,k}-\mu_i)^2.
\label{eq:group_advantage_clean}
\end{equation}
We clip the centered response-level advantage:
\begin{equation}
A_{i,k}
=
\operatorname{clip}
\left(
A^{\mathrm{resp}}_{i,k},
-c,\,
c
\right),
\label{eq:final_advantage_clean}
\end{equation}
Prompt-level scores such as $b(h_i)$ or trajectory-level annotations may be used to filter or stratify prompts before sampling, but they are not shifted into $A_{i,k}$; this keeps the advantages centered within each prompt group and avoids introducing a constant-gradient bias.

Policy updates use a clipped policy-gradient objective regularized by a reference model. Let
\begin{equation}
\rho_{i,k,t}
=
\frac{
\pi_{\phi}(a_{i,k,t} \mid p_i, a_{i,k,<t})
}{
\pi_{\mathrm{old}}(a_{i,k,t} \mid p_i, a_{i,k,<t})
}.
\end{equation}
We optimize
\begin{equation}
\mathcal{L}_{\mathrm{RL}}
=
-
\mathbb{E}
\left[
\min\!\Big(
\rho A,\,
\operatorname{clip}(\rho,1-\epsilon,1+\epsilon)A
\Big)
\right]
+
\eta \,\mathcal{D}_{\mathrm{ref}}(\pi_{\phi}, \pi_{\mathrm{ref}}),
\label{eq:rl_objective_clean}
\end{equation}
where $\mathcal{D}_{\mathrm{ref}}$ is a reference-model regularizer. Since the reward used in Eq.~\eqref{eq:rl_objective_clean} is computed by the frozen mental-state and reward model, the policy is trained \emph{under} the learned Theory-of-Mind model, even though the latent state is not fed explicitly to the policy at deployment.
The implementation supports gradient accumulation, reward clipping, cosine learning-rate scheduling, and checkpoint selection based on rolling validation reward.

\subsubsection{Deployment and Evaluation}
\label{app:deployment_evaluation}

At deployment time, only the learned policy is retained: the reward model is not used for test-time search or reranking. For open-ended interaction spaces, the policy directly generates a natural-language response. For structured action spaces, the same wrapper prompts the policy to emit a format-constrained output and parses it into the benchmark's expected action object. Episodes are then executed in each benchmark's native environment with the trained policy occupying one agent slot and the fixed partner or evaluator specified by the benchmark protocol.

\subsection{Dataset-Specific Instantiations}
\label{app:dataset_instantiations}

The shared objective is defined once in Appendix~\ref{app:shared_objective}. To separate methodological choices from dataset bookkeeping, each subsection below follows the same order: data construction, training instance, and policy/evaluation protocol. We specify only the context, utility target, mental-state supervision, and contrastive signal that change across datasets rather than re-expanding the same loss terms.

\subsubsection{\textsc{BigToM}}
\paragraph{Original dataset.}
\textsc{BigToM}~\citep{gandhi2023understanding} is a procedurally generated theory-of-mind benchmark of 200 scenarios. Each scenario is a five-sentence story with context, desire, perception cue, belief, and an external causal event that changes the object's state, paired with three questions (belief, desire, action) and gold answers for both branches: the agent \emph{is aware} of the causal event, or \emph{is not aware}.

\paragraph{Augmentation.}
We expand the seed set 25$\times$ by prompting Qwen3-235B with $k{=}4$ randomly sampled seeds as in-context exemplars and the original 19-field \textsc{BigToM} schema. We sample 5{,}000 new scenarios at
temperature~$1.0$ and retain only those passing a schema validator: exactly five sentences, the fourth beginning with a belief statement, distinct aware/not-aware branches, and a random event disjoint from the causal event. Generated scenarios are deduplicated against prior outputs and preserved the original \textsc{BigToM} schema, allowing drop-in compatibility with downstream tooling.

Because \textsc{BigToM} labels only first-order beliefs, we enrich each scenario with a second-order belief by prompting Qwen3-235B in reasoning mode (temperature~$0.4$) to produce: (i) an invented \emph{observer}, a plausible second character physically present at the event; (ii) a paraphrase of the story; and (iii) for each of the aware and not-aware conditions, a first-order belief (the agent's
belief about the object), a second-order belief (the agent's belief about what the observer believes about the object), and a 2--4 sentence chain-of-thought rationale ending in the predicted action. A
validator rejects outputs with empty fields or identical first-order beliefs across conditions.

\paragraph{Resulting training set.}
Each annotated synthetic scenario is flattened into two training examples, one per perceptual condition, yielding 9{,}964 rows from 5K generated \textsc{BigToM}-style scenarios. Every row carries \textsc{BigToM}-compatible gold labels (\texttt{gold\_belief}, \texttt{gold\_action}) alongside our added fields (\texttt{first\_order\_belief}, \texttt{second\_order\_belief}, \texttt{rationale}, \texttt{observer\_name}, \texttt{observer\_role}, \texttt{paraphrase}). This synthetic set is used for training; the official \textsc{BigToM} benchmark and \textsc{TOMi} are held out for evaluation. The rationale serves as the supervision target for mental-state reasoning, and the second-order belief enables training and evaluation of higher-order theory-of-mind beyond the scope of the original benchmark.

\paragraph{Training instance.}
For a scenario $s$, perceptual branch $b\in\{\mathrm{aware},\mathrm{not\_aware}\}$, initial-belief flag $\iota$, and question $q$, the context is $c_i^{\mathrm{BT}}=[\operatorname{story}_{\iota}(s);q]$. The positive answer $a_i^{+}=y(s,b,q)$ is paired with the opposite-branch answer $a_i^{-}=y(s,\bar b,q)$ as a hard negative. The scalar utility target $u_i^{\mathrm{BT}}\in\{0,1\}$ denotes branch-consistent correctness, so this setting uses scalar reward regression and an unmasked preference term. Generated first- and second-order beliefs supervise mental-state reconstruction, while the rationale is included in the mental-reasoning target. The branch label is used only by the optional aware/not-aware diagnostic classifier described above.

\paragraph{Optimization and evaluation.}
Policy learning uses mental-prefix SFT on the same QA prompts followed by GRPO, with the frozen scalar action-conditioned utility head providing the response reward and a reference-policy KL penalty. Prompt-level latent-only predictions are retained for diagnostics rather than added to the group-relative advantage. Evaluation uses the official true-belief, false-belief, and paired TB$\land$FB metrics.

\subsubsection{\textsc{Sotopia}}

\paragraph{Original dataset.}
\textsc{Sotopia}-$\pi$~\citep{WangEtAl2024SOTOPIAPi} builds on the \textsc{Sotopia} environment~\citep{zhou2025sotopia_s4}, which consists of 90 social scenarios and 40 character profiles. Each episode pairs two characters in a scenario and assigns each a private social goal, covering negotiation, collaboration, and competition. Character profiles include name, gender, occupation, personality description, and secrets. The released dataset contains multi-turn social interaction episodes, each with a full conversation and episode-level scores across seven social dimensions: believability, relationship, knowledge, secret-keeping, social rules, financial benefits, and goal completion.

\paragraph{Augmentation.}
We apply a two-stage GPT-4o annotation pipeline to each episode, decomposing each episode turn by turn and annotating every turn for both agents:

\textbf{Stage A}: reward scores with reasoning across all seven social dimensions, and first-order mental state annotations -- \texttt{partner\_belief} (what the agent believes about the partner's hidden intentions), \texttt{strategic\_intent} (the agent's sub-goal for this specific turn), \texttt{thought\_process} (the latent reasoning connecting belief to action), and \texttt{hard\_negative\_response} (a fluent but socially poor alternative response for contrastive learning).

\textbf{Stage B}: adds second-order theory-of-mind annotations for each agent-turn -- \texttt{second\_order\_belief} (what agent B thinks agent A believes), \texttt{second\_order\_intent} (what agent B thinks agent A intends), and \texttt{second\_order\_thought} (what agent B thinks agent A is thinking).

\paragraph{Resulting training set.}
In addition to the original \textsc{Sotopia}-$\pi$ conversation episode context (scenario, agent profiles, goals, secrets, and conversation), we generated per-turn reward scores with reasoning across all seven dimensions for both agents, alongside first-order and second-order mental state annotations, yielding 1,647 fully annotated episodes. These augmented \textsc{Sotopia}-$\pi$ episodes are used for training the mental-state/reward model; official \textsc{Sotopia} evaluation episodes are held out.

\paragraph{Training instance.}
For a \textsc{Sotopia} turn $t$ in episode $e$, the context includes the public scenario, both character profiles, private goals/secrets for the acting agent, and the dialogue prefix:
\begin{equation}
c_{e,t}^{\mathrm{ST}}
=
\big[
s_e;\ \rho_A;\rho_B;\ g_A;g_B;\ \operatorname{secret}_A;\operatorname{secret}_B;\ h_{e,<t}
\big].
\label{eq:sotopia_context}
\end{equation}
The positive action is the observed socially grounded utterance $a_{e,t}^{+}$, while $a_{e,t}^{-}$ is a generated hard-negative response that remains fluent but violates a social, strategic, or private-goal constraint. The utility vector is seven-dimensional,
\begin{equation}
\mathbf{u}_{e,t}^{\mathrm{ST}}
=
\big[
u_{\mathrm{bel}},u_{\mathrm{rel}},u_{\mathrm{know}},u_{\mathrm{secret}},
u_{\mathrm{rules}},u_{\mathrm{fin}},u_{\mathrm{goal}}
\big]\in[0,1]^7,
\label{eq:sotopia_utility}
\end{equation}
obtained by normalizing the \textsc{Sotopia} dimensions from their official score ranges. The first-order mental target concatenates partner belief, strategic intent, and thought process; the second-order target concatenates second-order belief, second-order intent, and second-order thought.

\paragraph{Optimization and evaluation.}
\textsc{Sotopia} instantiates all four supervised terms in Eq.~\eqref{eq:stage1_objective_expanded}: utility regression fits the seven-dimensional score, mental reconstruction decodes the two concatenated ToM targets, rationale conditioning predicts the same utility vector, and preference learning contrasts the observed response with the generated negative. The mask $\gamma_{e,t,d}^{\mathrm{ST}}$ activates the preference penalty only for dimensions intentionally degraded by that negative, avoiding penalties on qualities such as fluency or believability when they remain valid. Latent-only utility and future-action prediction are used only as the optional regularizers summarized above. GRPO scalarizes the seven predicted dimensions with $w_{\mathrm{ST}}\in\Delta^{6}$, and the deployed policy is evaluated with the official \textsc{Sotopia} evaluator on those dimensions and their mean.

\subsubsection{\textsc{MMRole}}

\paragraph{Original dataset.}
\textsc{MMRole}~\citep{dai2025mmrole} is a multimodal role-playing benchmark consisting of 85 characters, 11K images, and 14K single or multi-turn dialogues, yielding 85K training samples. Characters span fictional figures (Marvel heroes, Harry Potter, Friends), real historical persons (Kobe Bryant, Leonardo da Vinci), and Chinese historical figures. Each episode pairs a character profile (including personality, life story, relationships, and catchphrases) with a real image from COCO or movie/TV scripts. Three dialogue types are covered: \emph{comment} (a character comments on an image), \emph{human-role} (a human and a character discuss an image), and \emph{inter-role} (two characters interact about an image). The original dataset evaluates role-playing quality across eight dimensions: coherency, fluency, image-text relevance, instruction adherence, knowledge consistency, personality consistency, response accuracy, and tone consistency.

\paragraph{Augmentation.} We focus exclusively on \emph{inter-role} dialogues and apply a three-step
pipeline to add visual theory-of-mind annotations.

\textbf{Step 1 -- Per-turn restructuring}: Each inter-role dialogue is decomposed into per-turn snapshots. Each turn contains the speaker and partner profiles, the image, the full dialogue history up to that turn, and the current utterance, producing 36,744 per-turn examples.

\textbf{Step 2 -- Visual ToM annotation}: Each per-turn example is annotated by GPT-4o with access to the actual image, producing structured belief states:
\begin{itemize}
    \item \textbf{Visual percepts}: objects in the scene and their salience (high/medium/low/none) for each character \emph{based on their personality} --- a nurse and a politician attend to different elements of the same image.

    \item \textbf{First-order beliefs}: what the speaker believes the partner is visually attending to and intending.

    \item \textbf{Second-order beliefs}: what the speaker believes the partner believes about the speaker.

    \item \textbf{Belief probes}: structured question-answer pairs across five types (visual percept, first-order belief, second-order belief, intent inference, false belief) at three difficulty levels.

    \item \textbf{Contrastive belief states}: a ToM-violating alternative response for preference learning.
\end{itemize}

\textbf{Step 3 -- Validation}: Annotations are quality-filtered across six
checks --- object count, salience asymmetry between agents, belief-scene
grounding, probe answer distinctiveness, contrastive response distinctiveness,
and second-order belief consistency --- producing a clean set of 15,607
annotated turns.

\paragraph{Resulting training set.}
The 15,607 annotated turns are converted into four training formats using a
character-stratified split (74 train characters, 14 held-out test characters):
\begin{itemize}
    \item \textbf{Belief prediction} (12,762 train rows): given image, agent
    profiles, and dialogue history, predict the full structured belief state.
    \item \textbf{Preference pairs} (50,952 train rows): ToM-aligned responses paired against four types of violation, wrong visual attribution, wrong first-order belief, wrong second-order belief, and no theory-of-mind baseline, yielding four contrastive pairs per example.
    \item \textbf{Probe QA} (63,810 train rows): direct question-answer pairs
    for ToM evaluation and reward model fine-tuning.
    \item \textbf{Salience prediction} (25,524 train rows): given image and
    one character profile, predict per-object salience ratings.
\end{itemize}
Thus, \textsc{MMRole} uses the released training dialogues as the base resource, augments them with visual first-order/second-order mental labels and contrastive ToM violations, and evaluates on the held-out role split under the official \textsc{MMRole} protocol.

\paragraph{Training instance.}
For \textsc{MMRole}, the context is multimodal:
\begin{equation}
c_i^{\mathrm{MM}}
=
\big[
I_i;\ \rho_i^{\mathrm{speaker}};\rho_i^{\mathrm{partner}};\
h_{i,<t};\ \operatorname{role\_instruction}_i
\big],
\label{eq:mmrole_context}
\end{equation}
where $I_i$ is the image and $\rho$ denotes the character profile. A vision-language backbone jointly encodes the image and textual context, and its pooled representation parameterizes the two latent posteriors.
The positive response is the current in-character utterance, while negatives are contrastive ToM violations that alter visual grounding, first-order belief, second-order belief, or remove ToM reasoning. For each negative, $\gamma_{i,d}^{\mathrm{MM}}$ marks which official dimensions the contrastive edit is intended to violate. We obtain $\mathbf{u}_{i,-}^{\mathrm{MM}}$ by applying the same official \textsc{MMRole} evaluator and normalization to the contrastive negative response under the unchanged image, profile, and dialogue context. If a particular ablation does not score negatives with the evaluator, this regression term is masked. The joint reward vector is the official eight-dimensional \textsc{MMRole} score:
\begin{equation}
\mathbf{u}_i^{\mathrm{MM}}
=
\big[
u_{\mathrm{IA}},u_{\mathrm{Flu}},u_{\mathrm{Coh}},u_{\mathrm{ITR}},
u_{\mathrm{RA}},u_{\mathrm{PC}},u_{\mathrm{KC}},u_{\mathrm{TC}}
\big]\in[0,1]^8 .
\label{eq:mmrole_reward_vec}
\end{equation}
For diagnostics, we also derive an optional three-dimensional latent-only target covering first-order belief, second-order belief, and belief divergence; this is one of the lightweight probe regressors summarized in Appendix~\ref{app:shared_objective} rather than a core term.

\paragraph{Optimization and evaluation.}
The multimodal reward model uses the shared utility-regression, preference, mental-reconstruction, and KL terms. Utility regression fits the official eight-dimensional score, visual mental reconstruction decodes first- and second-order targets conditioned on the image, and $\gamma_{i,d}^{\mathrm{MM}}$ restricts each preference comparison to dimensions changed by that contrastive edit. When an official score for a negative response is available, we optionally regress to it directly; the three-dimensional latent-only ToM target and future-action prediction are likewise auxiliary regularizers rather than defining components of the method.

Policy training then uses visual mental-prefix SFT, GRPO with the frozen multimodal reward vector scalarized by $w_{\mathrm{MM}}\in\Delta^{7}$, and standard DPO on the contrastive preference pairs. Evaluation follows the official \textsc{MMRole} protocol and reports the normalized eight dimensions and their overall score.

Appendix Table~\ref{tab:dataset-comparison} summarizes which supervision fields are present in the original releases versus our augmented training resources. In terms of scale, our training data consists of 1,647 annotated \textsc{Sotopia}-$\pi$ episodes, 5K synthetic \textsc{BigToM}-style scenarios flattened to 9,964 rows, and 15,607 \textsc{MMRole} inter-role turns converted into belief-prediction, preference, probe-QA, and salience-training formats. Official \textsc{BigToM}, \textsc{TOMi}, \textsc{Sotopia}, \textsc{MMRole} held-out roles, \textsc{Craigslist-bargain}, and \textsc{FANToM} are used as evaluation or transfer settings rather than as training data.

\newcommand{\cmark}{\textcolor{green!60!black}{\ding{51}}}
\newcommand{\xmark}{\textcolor{red!80!black}{\ding{55}}}
\begin{table*}[h]
\centering
\caption{Comparison of original and augmented datasets across annotation dimensions.}
\label{tab:dataset-comparison}
\setlength{\tabcolsep}{2pt}
\renewcommand{\arraystretch}{1.2}
\resizebox{\textwidth}{!}{%
\begin{tabular}{lcccccccc}
\toprule
\textbf{Dataset}
  & \shortstack{\textbf{Multi-}\\\textbf{modal}}
  & \shortstack{\textbf{Episode}\\\textbf{Reward}}
  & \shortstack{\textbf{Per-turn}\\\textbf{Reward}}
  & \shortstack{\textbf{Intent}\\\textbf{Annot.}}
  & \shortstack{\textbf{1st-order}\\\textbf{Belief}}
  & \shortstack{\textbf{2nd-order}\\\textbf{Belief}}
  & \shortstack{\textbf{Rationale}\\\textbf{/ CoT}}
  & \shortstack{\textbf{Hard}\\\textbf{Negatives}} \\
\midrule
\textsc{BigToM} (original)        & \xmark & \xmark & \xmark & \xmark & \cmark & \xmark & \xmark & \xmark \\
\textsc{BigToM} (ours)            & \xmark & \xmark & \xmark & \cmark & \cmark & \cmark & \cmark & \cmark \\
\midrule
\textsc{Sotopia}-$\pi$ (original) & \xmark & \cmark & \xmark & \xmark & \xmark & \xmark & \xmark & \xmark \\
\textsc{Sotopia}-$\pi$ (ours)     & \xmark & \cmark & \cmark & \cmark & \cmark & \cmark & \cmark & \cmark \\
\midrule
\textsc{MMRole} (original)        & \cmark & \cmark & \xmark & \xmark & \xmark & \xmark & \xmark & \xmark \\
\textsc{MMRole} (ours)            & \cmark & \cmark & \cmark & \cmark & \cmark & \cmark & \xmark & \cmark \\
\bottomrule
\end{tabular}
}
\end{table*}

\subsubsection{Mental-State Annotation and Human Verification}
\label{app:annotation-validation}

Mental-state annotations are generated from the source evidence available in each benchmark, including the scenario or image, character profiles, dialogue history, and current turn. Before training, every supervision set passes automated structural and semantic-consistency checks. On \textsc{MMRole}, the six-check validator retained $15{,}607/15{,}919$ training turns ($98.04\%$) and $291/294$ examples from the official test partition.

We additionally conducted a blinded human audit of $100$ randomly sampled annotations ($50$ each from \textsc{MMRole} and \textsc{Sotopia}). Twelve evaluators saw the underlying evidence and generated annotation, but not the dataset or generator identity, and rated whether each mental-state claim was accurate and supported on a five-point scale. The overall mean was $\mathbf{4.45/5}$ ($89.0\%$ after linear rescaling), with $90/100$ annotations receiving at least $4$ and $56$ receiving $5$. The lower-rated cases primarily involved overconfident or unsupported inferences, underspecified claims, or contradictions with the dialogue or image. Because the audit prioritizes coverage of distinct records rather than repeated ratings per record, we treat it as a blinded quality spot-check rather than an inter-rater agreement study.

\begin{table}[h]
\centering
\caption{Blinded human audit of generated mental-state annotations. Faithfulness linearly rescales the mean rating to $0$--$100\%$.}
\label{tab:human-audit}
\small
\setlength{\tabcolsep}{4pt}
\begin{tabular}{lccc}
\toprule
Subset & $n$ & Mean ($1$--$5$) & Faithfulness (\%) \\
\midrule
\textsc{MMRole} & $50$ & $4.30$ & $86.0$ \\
\textsc{Sotopia} & $50$ & $4.60$ & $92.0$ \\
\midrule
\textbf{Overall} & $\mathbf{100}$ & $\mathbf{4.45}$ & $\mathbf{89.0}$ \\
\bottomrule
\end{tabular}
\end{table}

\subsection{Additional Results and Ablations}
\label{app:additional_results_ablations}

\paragraph{Zero-Shot Multi-Party Transfer to \textsc{FANToM}.}
\label{app:fantom-transfer}
We test whether a mental-model policy trained only on \textsc{BigToM}-style data transfers beyond dyadic interaction. Without any \textsc{FANToM} training, fine-tuning, or adaptation, we evaluate Base, SFT, and our policy on the first $5{,}000$ flattened probes of \textsc{FANToM}~\citep{kim2023fantom}, a multi-party, information-asymmetric ToM benchmark, using its official evaluation protocol and scorer. Our policy improves over Base on all four metrics and yields the strongest second-order score ($52.2$ vs.\ $49.7$ for Base and $41.4$ for SFT), providing preliminary evidence that the learned mental-model signal transfers to multi-party reasoning.

\begin{table*}[t]
\centering
\caption{Zero-shot \textsc{BigToM}$\rightarrow$\textsc{FANToM} transfer on $5{,}000$ probes using the official scorer. Scores are percentages; higher is better.}
\label{tab:fantom-transfer}
\small
\setlength{\tabcolsep}{5pt}
\begin{tabular}{lcccc}
\toprule
Model & Belief choice & Belief distance & First order & Second order \\
\midrule
Base & $25.4$ & $35.2$ & $26.8$ & $49.7$ \\
SFT & $26.5$ & $33.8$ & $29.4$ & $41.4$ \\
\textbf{Ours} & $\mathbf{27.9}$ & $\mathbf{38.0}$ & $\mathbf{29.7}$ & $\mathbf{52.2}$ \\
\bottomrule
\end{tabular}
\end{table*}

\begin{figure}[h]
\centering
\begin{minipage}[c]{0.65\textwidth}
  \centering
  \includegraphics[width=\linewidth]{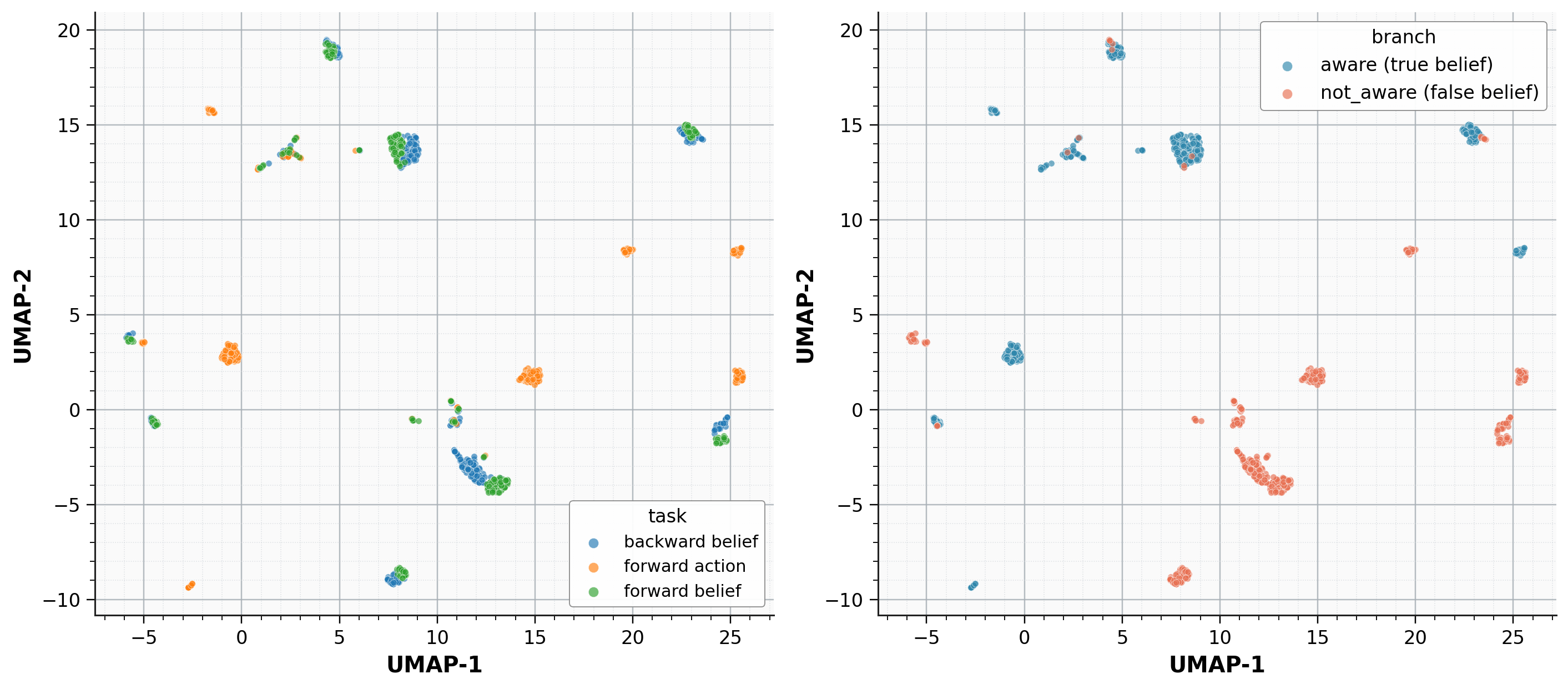}
\end{minipage}\hfill
\begin{minipage}[c]{0.32\textwidth}
  \caption{Mental-latent geometry from an unsupervised UMAP of latent $z$. \textbf{Left:} coloring by task shows task/scenario macro-clusters. \textbf{Right:} coloring by branch shows aware and not-aware samples remain near paired scenarios but are displaced within task regions, consistent with branch F1 $=0.989$.}
  \label{fig:latent-umap}
\end{minipage}
\end{figure}

\paragraph{\textsc{BigToM} Latent Geometry.}
The unsupervised UMAP in Figure~\ref{fig:latent-umap} provides a view of the \textsc{BigToM} latent geometry. Task identity forms the dominant macro-structure, while aware and not-aware branches remain locally separated inside nearby task/scenario regions. This supports the probe results in the main analysis: belief state is not the largest visual axis of variation, but it is still represented in a recoverable and geometrically coherent direction.

\begin{table*}[h]
\centering
\caption{
\textsc{BigToM} official-protocol results.
TB denotes true-belief accuracy, FB denotes false-belief accuracy, and TB$\land$FB denotes paired accuracy, where both the true-belief and false-belief versions of the same scenario must be answered correctly.
}
\label{tab:bigtom-main-supp}
\small
\setlength{\tabcolsep}{5pt}
\begin{tabular}{llcccccc}
\toprule
& & \multicolumn{3}{c}{Without Initial Belief} & \multicolumn{3}{c}{With Initial Belief} \\
\cmidrule(lr){3-5} \cmidrule(lr){6-8}
{Task} & {Model} & TB & FB & TB$\land$FB & TB & FB & TB$\land$FB \\
\midrule

\multirow{7}{*}{Forward Belief}
& GPT-4 (0-shot)      & 99.0 & 98.0 & 97.0 & 91.0 & 99.0 & 90.0 \\
& GPT-4 (0-shot-CoT)  & 99.0 & 99.0 & 98.0 & 99.0 & 99.0 & 98.0 \\
& GPT-4 (1-shot)      & 99.0 & 99.0 & 97.0 & 97.0 & 99.0 & 96.0 \\
& GPT-4 (1-shot-CoT)  & 100.0 & 99.0 & 99.0 & 97.0 & 99.0 & 96.0 \\
& Base                & 83.0 & 94.0 & 77.0 & 49.5 & 98.5 & 48.0 \\
& SFT      & 95.5 & 88.5 & 84.5 & 94.0 & 96.5 & 92.0 \\
& Ours                & \textbf{97.0} & \textbf{98.0} & \textbf{96.0} & \textbf{97.5} & \textbf{99.5} & \textbf{97.5} \\
\midrule

\multirow{7}{*}{Forward Action}
& GPT-4 (0-shot)      & 98.0 & 81.0 & 79.0 & 98.0 & 92.0 & 90.0 \\
& GPT-4 (0-shot-CoT)  & 99.0 & 88.0 & 87.0 & 99.0 & 96.0 & 95.0 \\
& GPT-4 (1-shot)      & 100.0 & 98.0 & 98.0 & 100.0 & 100.0 & 100.0 \\
& GPT-4 (1-shot-CoT)  & 100.0 & 100.0 & 100.0 & 100.0 & 99.0 & 99.0 \\
& Base                & 99.5 & 64.0 & 63.5 & \textbf{99.5} & 68.0 & 67.5 \\
& SFT      & 93.0 & 82.5 & 76.5 & 92.0 & 80.0 & 73.0 \\
& Ours                & \textbf{99.5} & \textbf{79.0} & \textbf{78.5} & 99.0 & \textbf{81.0} & \textbf{80.0} \\
\midrule

\multirow{7}{*}{Backward Belief}
& GPT-4 (0-shot)      & 86.0 & 53.0 & 40.0 & 62.0 & 77.0 & 40.0 \\
& GPT-4 (0-shot-CoT)  & 84.0 & 54.0 & 38.0 & 76.0 & 63.0 & 40.0 \\
& GPT-4 (1-shot)      & 68.0 & 85.0 & 53.0 & 57.0 & 92.0 & 49.0 \\
& GPT-4 (1-shot-CoT)  & 83.0 & 75.0 & 58.0 & 81.0 & 85.0 & 65.0 \\
& Base                & 85.0 & 92.5 & 77.5 & 53.5 & 99.0 & 52.5 \\
& SFT       & 94.5 & 90.5 & 85.0 & 94.5 & 95.5 & 91.0 \\
& Ours                & \textbf{97.0} & \textbf{98.5} & \textbf{95.5} & \textbf{98.5} & \textbf{99.5} & \textbf{98.5} \\
\bottomrule
\end{tabular}
\end{table*}

\paragraph{\textsc{BigToM} official-protocol comparison.}
Table~\ref{tab:bigtom-main-supp} expands the main \textsc{BigToM} result with the full official prompting protocol, including zero-shot, CoT, and one-shot GPT-4 references. The most diagnostic columns are TB$\land$FB, because they require the model to answer both true- and false-belief branches of the same scenario correctly rather than solving only the easier branch. Our model gives the strongest open-source paired accuracy in forward belief and backward belief, and improves the paired forward-action scores over the base model in both settings. This pattern suggests that the mental-state reward signal mainly helps with counterfactual consistency: the policy is less likely to collapse to the observable world state when the initial belief sentence or the false-belief branch changes.

\begin{figure}[H]
  \centering
  \begin{minipage}[c]{0.45\linewidth}
    \centering
    \includegraphics[width=\linewidth]{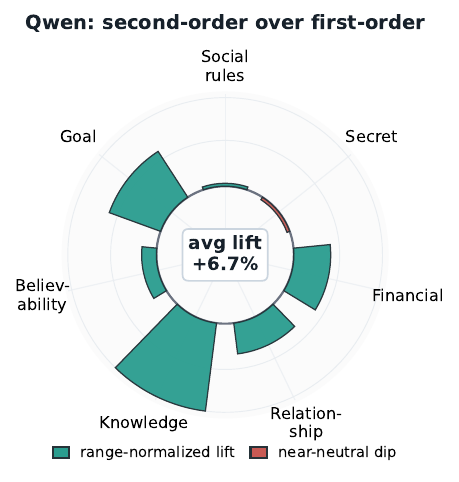}
  \end{minipage}\hfill
  \begin{minipage}[c]{0.51\linewidth}
    \caption{Range-normalized lift from Qwen2.5-7B (ours w/ 1st order) to Qwen2.5-7B (ours w/ 2nd order) on \textsc{Sotopia}. Petal lengths encode percentage-of-range changes for each official dimension rather than raw scores. The center reports the relative lift in the average score, highlighting that second-order ToM improves the aggregate profile while keeping the comparison scale-free.}
    \label{fig:qwen-order-lift}
\end{minipage}
\end{figure}

\paragraph{First- versus second-order mental state.}
Figure~\ref{fig:qwen-order-lift} visualizes the comparison of Qwen2.5-7B variant for first-order and second-order mental-state latents on the \textbf{\textsc{Sotopia}} benchmark. Each petal shows the second-order change as a percentage of the official range for that \textsc{Sotopia} dimension, while the center badge reports the relative lift in the average score. The second-order latent gives a broader outward profile, especially on task-facing dimensions such as knowledge, goal completion, financial reasoning, and relationship management, suggesting that recursive partner modeling adds useful signal beyond a first-order mental state alone.

\begin{figure}[H]
  \centering
  \begin{minipage}[c]{0.5\linewidth}
    \centering
    \includegraphics[width=\linewidth]{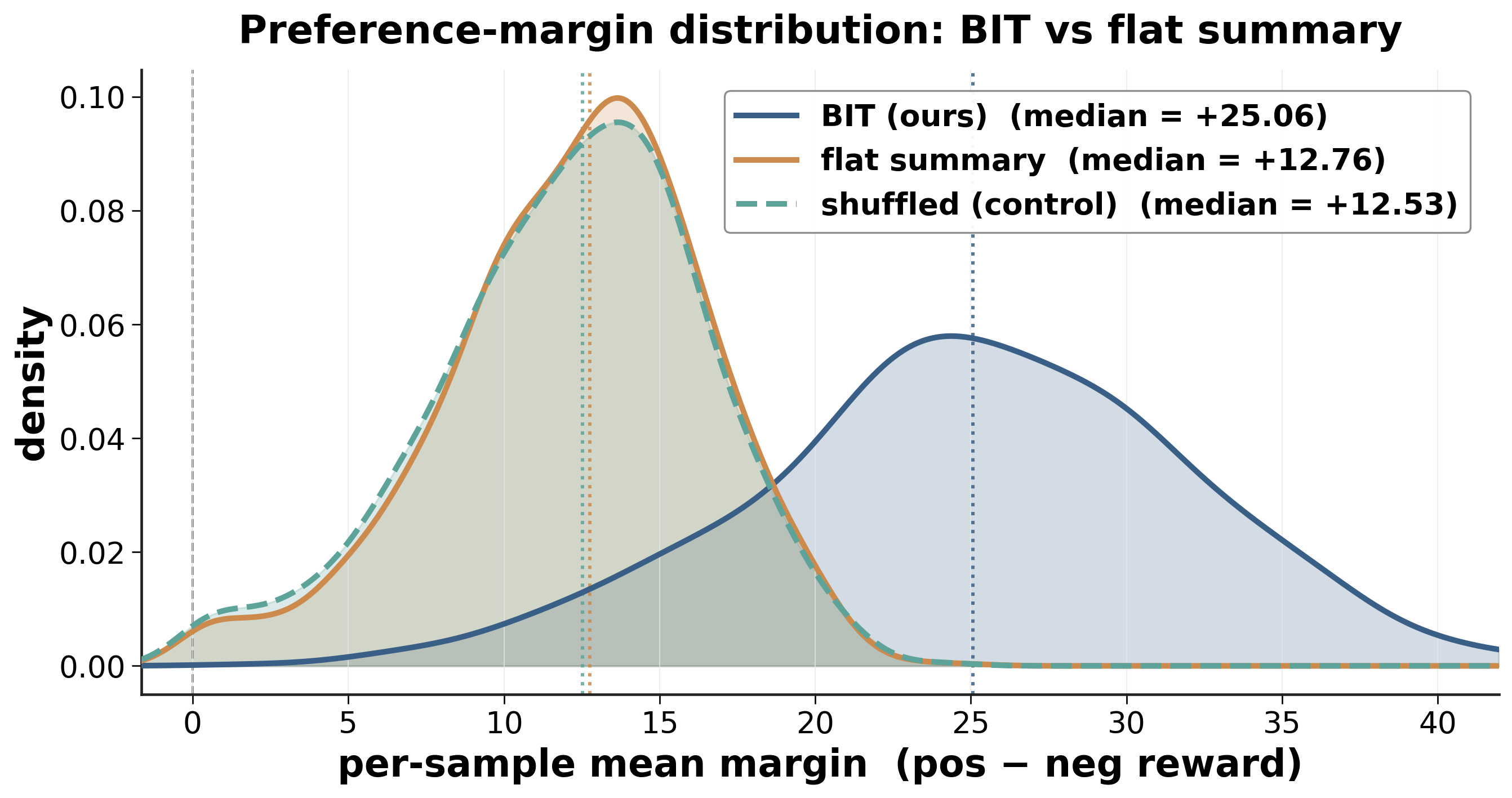}
  \end{minipage}\hfill
  \begin{minipage}[c]{0.5\linewidth}
    \caption{\textbf{Preference-margin distribution.} BIT decomposition produces sharper preference margins than flat or shuffled mental supervision. Kernel density of the per-sample margin $r^{+}-r^{-}$ on 1,508 \textsc{Sotopia} validation turns. Dotted vertical lines mark medians; the dashed line marks zero. BIT ranks positives over negatives with roughly $2{\times}$ higher confidence (median $+25.06$ vs. flat $+12.76$), and every BIT sample is correctly ranked with margin $\geq+1.40$. Flat overlaps the shuffled-label control, indicating that mental supervision improves calibration when belief, intent, and thought are routed to dedicated latent sub-blocks.}
    \label{fig:pref-margin-distribution}
  \end{minipage}
\end{figure}

\paragraph{BIT Factorization Sharpens Preference Margins.}
We ablate the BIT decomposition by retraining identical encoders with two control supervision signals: a \emph{flat summary}, where belief, intent, and thought are concatenated into a single free-form target decoded from the full latent, and a \emph{shuffled} variant, where the three labels are randomly reassigned to slots before training. All three variants reach near-ceiling preference accuracy on the \textsc{Sotopia} validation set. The preference loss, however, separates them by two orders of magnitude (BIT $9.6{\times}10^{-5}$ vs. flat $1.0{\times}10^{-2}$), because $\operatorname{softplus}(r^{-}-r^{+})$ measures the margin by which positives are ranked above negatives, not only whether the ranking is correct. Figure~\ref{fig:pref-margin-distribution} shows the per-sample margin $r^{+}-r^{-}$: BIT is centered at $+25.06$ median, with every sample correctly ranked at margin $\geq +1.40$, while flat summary is centered at $+12.76$ and overlaps the shuffled-label control ($+12.53$ median). This suggests that mental content alone is not enough; routing belief, intent, and thought to dedicated latent sub-blocks is what yields confident partner-conditioned preference separation.

\begin{figure}[H]
  \centering
  \begin{minipage}[c]{0.43\linewidth}
    \centering
    \includegraphics[width=\linewidth]{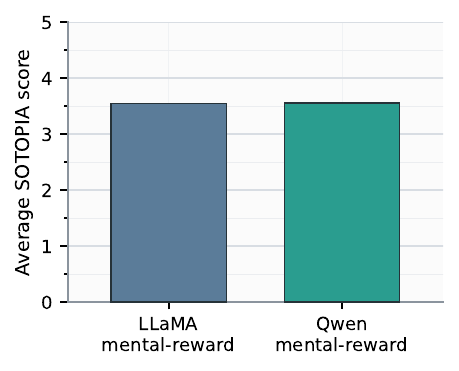}
  \end{minipage}\hfill
  \begin{minipage}[c]{0.53\linewidth}
    \caption{Cross-agent transfer of the mental-reward model on \textsc{Sotopia}. Bars compare the average scores of LLaMA2-7B (ours) with its same-architecture mental-reward model and LLaMA2-7B (ours w/ r-Qwen) with the Qwen mental-reward model. The transferred Qwen mental-reward model preserves overall performance, supporting its generalization across agents.}
    \label{fig:llama-rqwen-transfer}
\end{minipage}
\end{figure}

\paragraph{Cross-agent mental-reward transfer.}
Figure~\ref{fig:llama-rqwen-transfer} compares the aggregate \textsc{Sotopia} score under the two reward models. The transferred Qwen mental-reward model does not reduce performance and slightly improves the average score (3.557 vs. 3.541), indicating that the learned mental-reward architecture is not tied to a single policy family and can provide a useful scoring signal for another agent.

\begin{figure}[H]
  \centering
  \begin{minipage}[c]{0.61\linewidth}
    \centering
    \includegraphics[width=\linewidth]{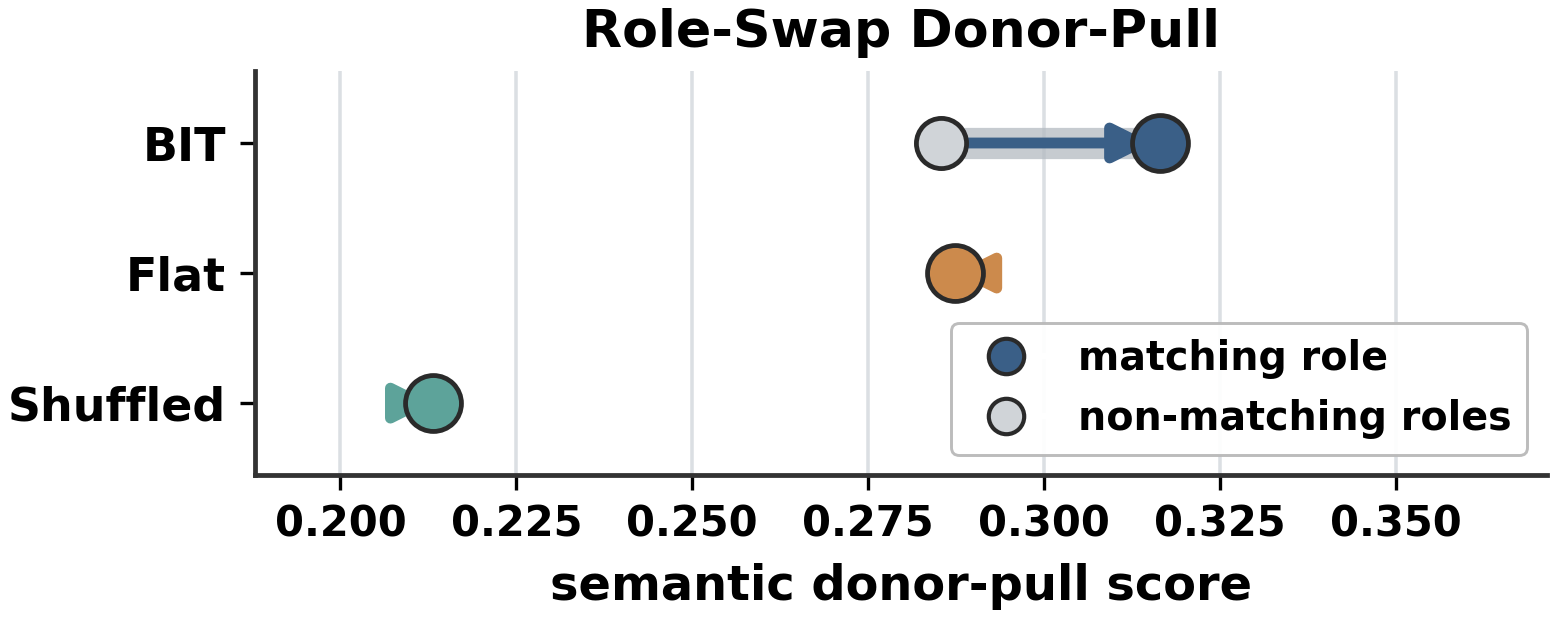}
  \end{minipage}\hfill
  \begin{minipage}[c]{0.35\linewidth}
    \caption{\textbf{Role-swap latent intervention.} Swapping a BIT role-specific block moves the corresponding role prediction more than unrelated roles. Flat and Shuffled variants do not show this separation, yielding more diffuse effects.}
    \label{fig:role-swap}
\end{minipage}
\end{figure}

\paragraph{Role-Addressable Latent Blocks.}
We additionally perform a role-swap intervention in latent space (Fig. \ref{fig:role-swap}). Although the absolute effects are small, BIT is the only variant where swapping a role-specific latent block produces a greater semantic pull on the matching role prediction than on non-matching roles; Flat and Shuffled remain diffuse. This suggests that the BIT decomposition induces weak but measurable role-addressability.

\begin{figure} [H]
  \centering
  \includegraphics[width=\linewidth]{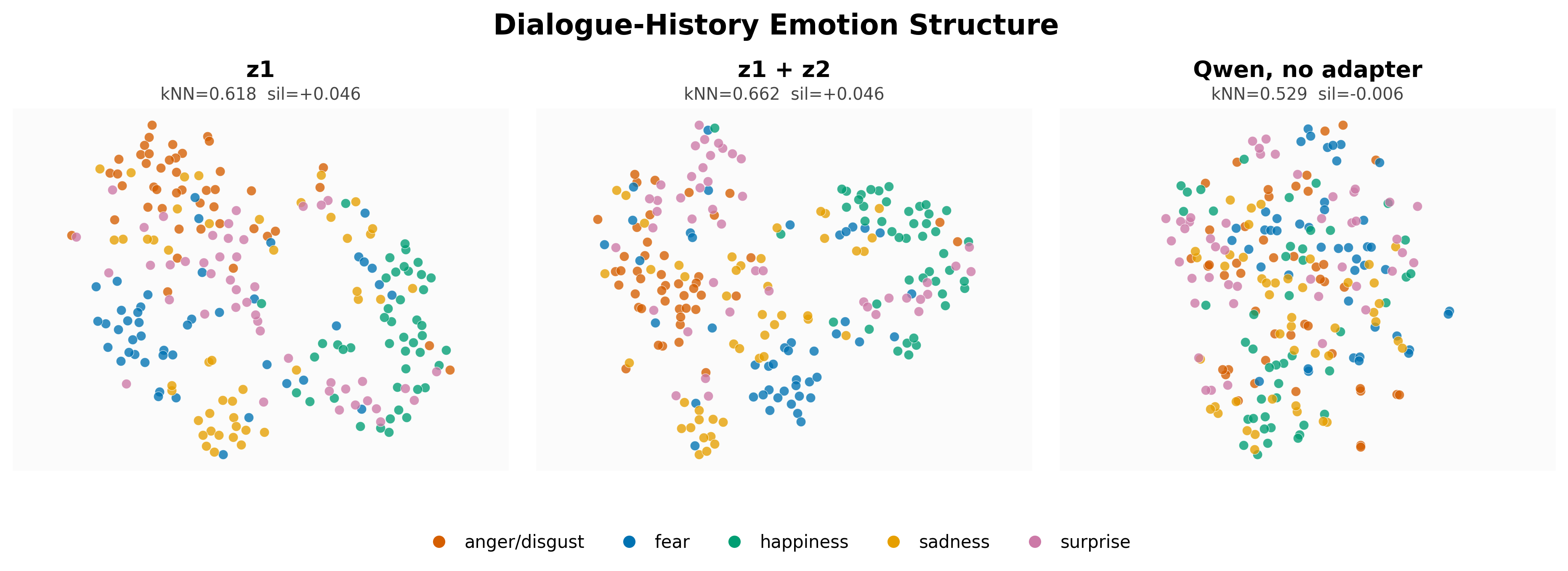}
  \caption{
  \textbf{t-SNE visualization} The concatenated latent representation [z1; z2] reaches 0.662 5-NN accuracy (silhouette +0.046) on the 5-way emotion task compared to 0.529 for Qwen without the adapter, and a chance baseline of 0.200. The +0.13 gap over the no-adapter baseline indicates that the learned mental latents encode emotion-relevant structure that is not present in the base Qwen representation. The corresponding t-SNE projection provides a qualitative visualization: classes appear more separated in the z1 and [z1; z2] projections, while the no-adapter Qwen projection shows more inter-class overlap.}
  \label{fig:emotion-latent-tsne}
\end{figure}

\paragraph{Interpreting z-space and Ablation on Emotion Structure in Dialogue-Level Mental Latents}
We conduct an ablation study to test whether the z-space (mental-state latent space) learned by the recursive ToM model captures affective information beyond the model's original \textsc{Sotopia} training objective. We use a dialogue-level DailyDialog~\cite{li2017dailydialogmanuallylabelledmultiturn} dataset subset filtered to contain only dialogues with a single consistent non-neutral emotion, ignoring neutral turns and merging anger/disgust due to their strong overlap. This yields a balanced 5-way emotion classification setting. Each dialogue is represented in a structured dialogue history format with turn indices and speaker markers, and we append the prompt suffix \textit{Emotion:} so that the final-token latent is aligned with an emotion-prediction query. We compare z-probe representations against no-adapter Qwen baseline, to test whether emotion separability comes from the learned mental latents rather than from the base language model alone. We assess the embeddings along two complementary axes:
\begin{enumerate}
    \item 5-fold cross-validated 5-NN accuracy and silhouette score, both computed on the standardized high-dimensional embeddings. The 5-NN accuracy measures how easily emotion labels can be inferred from the embedding geometry, while the silhouette score measures class compactness and separation in the original high-dimensional space.
    \item t-SNE visualization at perplexity 15 Figure~\ref{fig:emotion-latent-tsne}, which projects each embedding to two dimensions for qualitative inspection of class structure. It provides a complementary geometric view of how the classes are arranged.
\end{enumerate}

\begin{figure}[H]
  \centering
  \begin{minipage}[c]{0.48\linewidth}
    \centering
    \includegraphics[width=\linewidth]{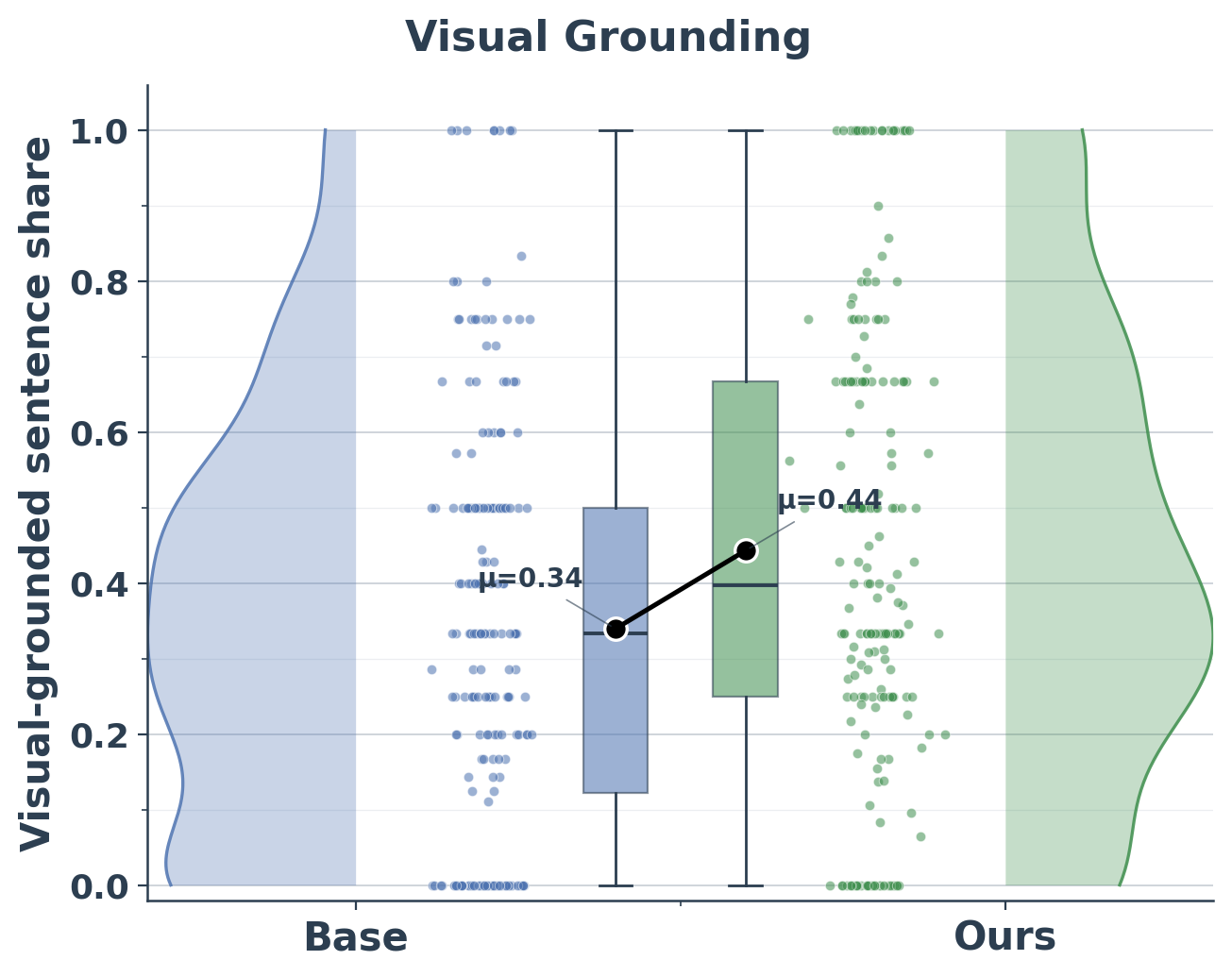}
  \end{minipage}\hfill
  \begin{minipage}[c]{0.48\linewidth}
    \caption{\textbf{Visual Grounding.} Visual grounding rate per response for Base vs. Ours on 200 held-out \textsc{MMRole} examples. Each point is one example's share of generated sentences that mention a gold scene object; black markers show means. The metric is computed using a token-overlap classifier between generated sentence tokens and the gold \texttt{scene\_objects} token bag, making it a proxy for grounding rather than grounding itself. Ours increases the rate by $\Delta=+0.103$ (paired Wilcoxon $p=1.3\times10^{-4}$, ***).}
    \label{fig:mmrole-visual-grounding}
  \end{minipage}
\end{figure}

\paragraph{Visual Grounding in \textsc{MMRole}.}
Figure~\ref{fig:mmrole-visual-grounding} provides an appendix-level check that the \textsc{MMRole} gains are accompanied by stronger use of visual evidence. We measure a response-level visual-grounding proxy on 200 held-out examples by splitting each generated response into sentences and computing the share of sentences that mention a gold scene object. The detector is an operational token-overlap classifier between generated sentence tokens and the gold \texttt{scene\_objects} token bag, so it should be read as a proxy for grounding rather than grounding itself. Under this metric, ours increases the mean visual grounding rate over the base model by $0.103$ with a paired Wilcoxon $p=1.3\times10^{-4}$, suggesting that mental-model training keeps generated role responses more tied to objects present in the scene.

\begin{figure}[H]
  \centering
  \begin{minipage}[c]{0.48\linewidth}
    \centering
    \includegraphics[width=\linewidth]{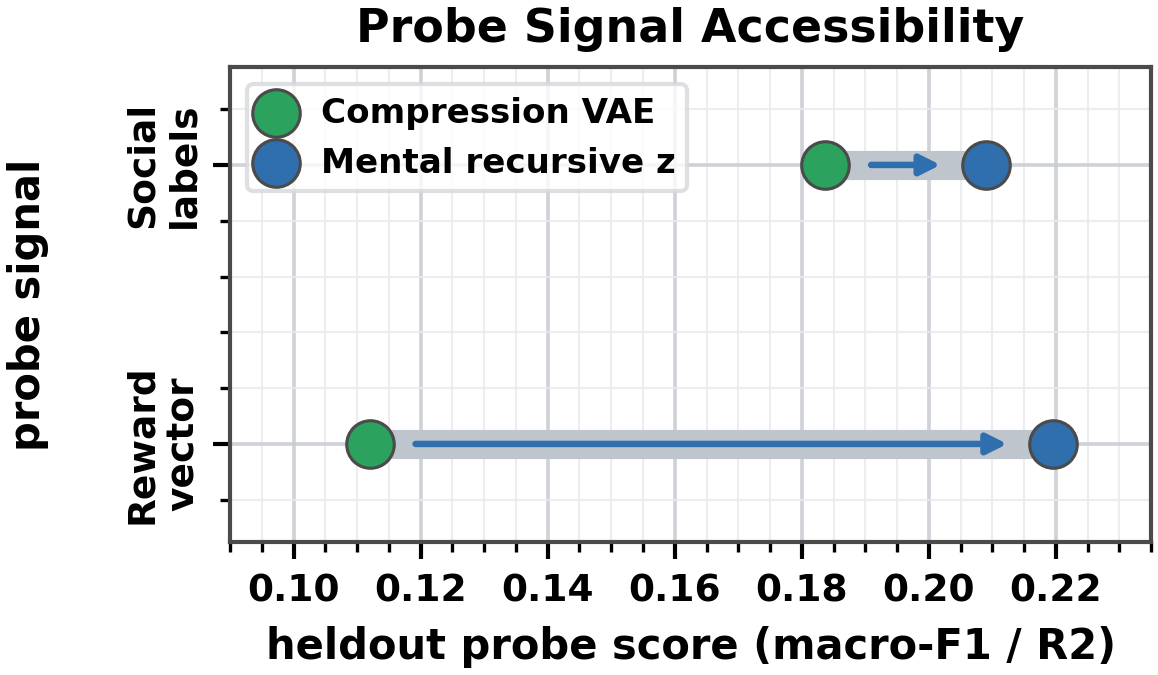}
  \end{minipage}\hfill
  \begin{minipage}[c]{0.48\linewidth}
    \caption{\textbf{Probe signal accessibility.} We compare the recursive mental-model latent against a compression-only VAE trained on the same \textsc{Sotopia}-$\pi$ contexts and responses with matched latent dimensionality. The VAE is optimized only for reconstruction, while the mental-model encoder is trained with ToM supervision and reward coupling. After freezing both encoders, we fit simple probes from $z$ and evaluate held-out recovery of social-label structure and reward-vector targets. Higher macro-F1/$R^2$ for the mental latent shows that downstream social and reward signals are more accessible than in a generic compression code.}
    \label{fig:task-signal-accessibility}
  \end{minipage}
\end{figure}

\paragraph{Mental-model is not a compression code. }
To test whether the mental-model latent is more than a compression code, we compare it against a compression-only VAE trained on the same \textsc{Sotopia}-$\pi$ contexts and responses, using the same latent dimensionality and a reconstruction objective. We freeze both encoders, train lightweight linear probes from the latent to task-relevant targets, and evaluate recovery of social-label structure and reward-vector dimensions on held-out examples. Figure~\ref{fig:task-signal-accessibility} shows that the probes recover both targets more effectively from the mental latent than from the compression VAE. This indicates that the recursive mental-model objective does not merely compress the context; it organizes the latent space around downstream social and reward signals.

\subsection{Impact Statement}
\label{app:impact-statement}

This work studies explicit mental-model representations for multi-agent systems. The main positive impact is to make agent behavior more interpretable and controllable in settings where an agent must reason about another agent's goals, beliefs, knowledge, or visual perspective. Such representations can support safer assistants, better negotiation and collaboration agents, and clearer diagnostic tools for evaluating whether a model is relying on task-relevant social and perceptual signals rather than surface imitation. Our augmented datasets also provide structured supervision for studying Theory-of-Mind behavior across language-only and multimodal benchmarks.

The same capability also carries risks. Models that infer or exploit another agent's latent state could be misused for manipulation, persuasion, privacy invasion, or profiling, especially if deployed in high-stakes human-facing settings. The inferred variables in our framework should therefore be treated as task-conditioned modeling constructs, not as ground-truth psychological states. They may reflect annotation bias, dataset artifacts, or stereotypes in the underlying models and data. For this reason, deployment should include consent-aware data collection, privacy filtering, bias audits, uncertainty reporting, and restrictions against using inferred mental states for sensitive decisions about individuals. Our experiments are limited to benchmark and synthetic/augmented training settings, and future work should study human oversight and safety constraints before using such systems in real-world multi-agent applications.

\clearpage
\section*{NeurIPS Paper Checklist}

\begin{enumerate}

\item {\bf Claims}
    \item[] Question: Do the main claims made in the abstract and introduction accurately reflect the paper's contributions and scope?
    \item[] Answer: \answerYes{} %
    \item[] Justification: Refer to section \ref{sec:experimentation}
    \item[] Guidelines:
    \begin{itemize}
        \item The answer \answerNA{} means that the abstract and introduction do not include the claims made in the paper.
        \item The abstract and/or introduction should clearly state the claims made, including the contributions made in the paper and important assumptions and limitations. A \answerNo{} or \answerNA{} answer to this question will not be perceived well by the reviewers. 
        \item The claims made should match theoretical and experimental results, and reflect how much the results can be expected to generalize to other settings. 
        \item It is fine to include aspirational goals as motivation as long as it is clear that these goals are not attained by the paper. 
    \end{itemize}

\item {\bf Limitations}
    \item[] Question: Does the paper discuss the limitations of the work performed by the authors?
    \item[] Answer: \answerYes{} %
    \item[] Justification: Refer to section \ref{sec:conclusion}
    \item[] Guidelines:
    \begin{itemize}
        \item The answer \answerNA{} means that the paper has no limitation while the answer \answerNo{} means that the paper has limitations, but those are not discussed in the paper. 
        \item The authors are encouraged to create a separate ``Limitations'' section in their paper.
        \item The paper should point out any strong assumptions and how robust the results are to violations of these assumptions (e.g., independence assumptions, noiseless settings, model well-specification, asymptotic approximations only holding locally). The authors should reflect on how these assumptions might be violated in practice and what the implications would be.
        \item The authors should reflect on the scope of the claims made, e.g., if the approach was only tested on a few datasets or with a few runs. In general, empirical results often depend on implicit assumptions, which should be articulated.
        \item The authors should reflect on the factors that influence the performance of the approach. For example, a facial recognition algorithm may perform poorly when image resolution is low or images are taken in low lighting. Or a speech-to-text system might not be used reliably to provide closed captions for online lectures because it fails to handle technical jargon.
        \item The authors should discuss the computational efficiency of the proposed algorithms and how they scale with dataset size.
        \item If applicable, the authors should discuss possible limitations of their approach to address problems of privacy and fairness.
        \item While the authors might fear that complete honesty about limitations might be used by reviewers as grounds for rejection, a worse outcome might be that reviewers discover limitations that aren't acknowledged in the paper. The authors should use their best judgment and recognize that individual actions in favor of transparency play an important role in developing norms that preserve the integrity of the community. Reviewers will be specifically instructed to not penalize honesty concerning limitations.
    \end{itemize}

\item {\bf Theory assumptions and proofs}
    \item[] Question: For each theoretical result, does the paper provide the full set of assumptions and a complete (and correct) proof?
    \item[] Answer: \answerNA{} %
    \item[] Justification: The paper does not include theoretical results
    \item[] Guidelines:
    \begin{itemize}
        \item The answer \answerNA{} means that the paper does not include theoretical results. 
        \item All the theorems, formulas, and proofs in the paper should be numbered and cross-referenced.
        \item All assumptions should be clearly stated or referenced in the statement of any theorems.
        \item The proofs can either appear in the main paper or the supplemental material, but if they appear in the supplemental material, the authors are encouraged to provide a short proof sketch to provide intuition. 
        \item Inversely, any informal proof provided in the core of the paper should be complemented by formal proofs provided in appendix or supplemental material.
        \item Theorems and Lemmas that the proof relies upon should be properly referenced. 
    \end{itemize}

    \item {\bf Experimental result reproducibility}
    \item[] Question: Does the paper fully disclose all the information needed to reproduce the main experimental results of the paper to the extent that it affects the main claims and/or conclusions of the paper (regardless of whether the code and data are provided or not)?
    \item[] Answer: \answerYes{} %
    \item[] Justification: Refer to sections \ref{sec:experimentation}, \ref{app:dataset_instantiations}, and \ref{app:implementation_details}
    \item[] Guidelines:
    \begin{itemize}
        \item The answer \answerNA{} means that the paper does not include experiments.
        \item If the paper includes experiments, a \answerNo{} answer to this question will not be perceived well by the reviewers: Making the paper reproducible is important, regardless of whether the code and data are provided or not.
        \item If the contribution is a dataset and\slash or model, the authors should describe the steps taken to make their results reproducible or verifiable. 
        \item Depending on the contribution, reproducibility can be accomplished in various ways. For example, if the contribution is a novel architecture, describing the architecture fully might suffice, or if the contribution is a specific model and empirical evaluation, it may be necessary to either make it possible for others to replicate the model with the same dataset, or provide access to the model. In general. releasing code and data is often one good way to accomplish this, but reproducibility can also be provided via detailed instructions for how to replicate the results, access to a hosted model (e.g., in the case of a large language model), releasing of a model checkpoint, or other means that are appropriate to the research performed.
        \item While NeurIPS does not require releasing code, the conference does require all submissions to provide some reasonable avenue for reproducibility, which may depend on the nature of the contribution. For example
        \begin{enumerate}
            \item If the contribution is primarily a new algorithm, the paper should make it clear how to reproduce that algorithm.
            \item If the contribution is primarily a new model architecture, the paper should describe the architecture clearly and fully.
            \item If the contribution is a new model (e.g., a large language model), then there should either be a way to access this model for reproducing the results or a way to reproduce the model (e.g., with an open-source dataset or instructions for how to construct the dataset).
            \item We recognize that reproducibility may be tricky in some cases, in which case authors are welcome to describe the particular way they provide for reproducibility. In the case of closed-source models, it may be that access to the model is limited in some way (e.g., to registered users), but it should be possible for other researchers to have some path to reproducing or verifying the results.
        \end{enumerate}
    \end{itemize}

\item {\bf Open access to data and code}
    \item[] Question: Does the paper provide open access to the data and code, with sufficient instructions to faithfully reproduce the main experimental results, as described in supplemental material?
    \item[] Answer: \answerNo{} %
    \item[] Justification: We will release the code, models and data upon acceptance.
    \item[] Guidelines:
    \begin{itemize}
        \item The answer \answerNA{} means that paper does not include experiments requiring code.
        \item Please see the NeurIPS code and data submission guidelines (\url{https://neurips.cc/public/guides/CodeSubmissionPolicy}) for more details.
        \item While we encourage the release of code and data, we understand that this might not be possible, so \answerNo{} is an acceptable answer. Papers cannot be rejected simply for not including code, unless this is central to the contribution (e.g., for a new open-source benchmark).
        \item The instructions should contain the exact command and environment needed to run to reproduce the results. See the NeurIPS code and data submission guidelines (\url{https://neurips.cc/public/guides/CodeSubmissionPolicy}) for more details.
        \item The authors should provide instructions on data access and preparation, including how to access the raw data, preprocessed data, intermediate data, and generated data, etc.
        \item The authors should provide scripts to reproduce all experimental results for the new proposed method and baselines. If only a subset of experiments are reproducible, they should state which ones are omitted from the script and why.
        \item At submission time, to preserve anonymity, the authors should release anonymized versions (if applicable).
        \item Providing as much information as possible in supplemental material (appended to the paper) is recommended, but including URLs to data and code is permitted.
    \end{itemize}

\item {\bf Experimental setting/details}
    \item[] Question: Does the paper specify all the training and test details (e.g., data splits, hyperparameters, how they were chosen, type of optimizer) necessary to understand the results?
    \item[] Answer: \answerYes{} %
    \item[] Justification: Refer to sections \ref{sec:experimentation},\ref{app:dataset_instantiations} and \ref{app:implementation_details}
    \item[] Guidelines:
    \begin{itemize}
        \item The answer \answerNA{} means that the paper does not include experiments.
        \item The experimental setting should be presented in the core of the paper to a level of detail that is necessary to appreciate the results and make sense of them.
        \item The full details can be provided either with the code, in appendix, or as supplemental material.
    \end{itemize}

\item {\bf Experiment statistical significance}
    \item[] Question: Does the paper report error bars suitably and correctly defined or other appropriate information about the statistical significance of the experiments?
\item[] Answer: \answerNo{}
\item[] Justification: The paper reports official benchmark scores under fixed evaluation protocols and includes a paired Wilcoxon significance test for the \textsc{MMRole} visual-grounding analysis in Appendix~\ref{fig:mmrole-visual-grounding}. For \textsc{Sotopia}, we repeated the main evaluations multiple times and observed nearly identical aggregate scores, suggesting low evaluation variance under the fixed protocol. However, we do not report formal run-to-run error bars or confidence intervals for all main trained-policy comparisons, since repeating full 7B-scale LLM/VLM training and closed-source evaluations across multiple seeds is computationally and API-cost intensive. We therefore treat the main benchmark tables as point estimates and support the claims through consistency across benchmarks, model families, held-out splits, repeated \textsc{Sotopia} evaluations, and ablations.

    \item[] Guidelines:
    \begin{itemize}
        \item The answer \answerNA{} means that the paper does not include experiments.
        \item The authors should answer \answerYes{} if the results are accompanied by error bars, confidence intervals, or statistical significance tests, at least for the experiments that support the main claims of the paper.
        \item The factors of variability that the error bars are capturing should be clearly stated (for example, train/test split, initialization, random drawing of some parameter, or overall run with given experimental conditions).
        \item The method for calculating the error bars should be explained (closed form formula, call to a library function, bootstrap, etc.)
        \item The assumptions made should be given (e.g., Normally distributed errors).
        \item It should be clear whether the error bar is the standard deviation or the standard error of the mean.
        \item It is OK to report 1-sigma error bars, but one should state it. The authors should preferably report a 2-sigma error bar than state that they have a 96\% CI, if the hypothesis of Normality of errors is not verified.
        \item For asymmetric distributions, the authors should be careful not to show in tables or figures symmetric error bars that would yield results that are out of range (e.g., negative error rates).
        \item If error bars are reported in tables or plots, the authors should explain in the text how they were calculated and reference the corresponding figures or tables in the text.
    \end{itemize}

\item {\bf Experiments compute resources}
    \item[] Question: For each experiment, does the paper provide sufficient information on the computer resources (type of compute workers, memory, time of execution) needed to reproduce the experiments?
    \item[] Answer: \answerYes{} %
    \item[] Justification: Refer to sections \ref{sec:experimentation},  \ref{app:implementation_details} and \ref{app:dataset_instantiations}
    \item[] Guidelines:
    \begin{itemize}
        \item The answer \answerNA{} means that the paper does not include experiments.
        \item The paper should indicate the type of compute workers CPU or GPU, internal cluster, or cloud provider, including relevant memory and storage.
        \item The paper should provide the amount of compute required for each of the individual experimental runs as well as estimate the total compute. 
        \item The paper should disclose whether the full research project required more compute than the experiments reported in the paper (e.g., preliminary or failed experiments that didn't make it into the paper). 
    \end{itemize}
    
\item {\bf Code of ethics}
    \item[] Question: Does the research conducted in the paper conform, in every respect, with the NeurIPS Code of Ethics \url{https://neurips.cc/public/EthicsGuidelines}?
    \item[] Answer: \answerYes{} %
    \item[] Justification: Yes paper conforms with NeurIPS code of ethics
    \item[] Guidelines:
    \begin{itemize}
        \item The answer \answerNA{} means that the authors have not reviewed the NeurIPS Code of Ethics.
        \item If the authors answer \answerNo, they should explain the special circumstances that require a deviation from the Code of Ethics.
        \item The authors should make sure to preserve anonymity (e.g., if there is a special consideration due to laws or regulations in their jurisdiction).
    \end{itemize}

\item {\bf Broader impacts}
    \item[] Question: Does the paper discuss both potential positive societal impacts and negative societal impacts of the work performed?
    \item[] Answer: \answerYes{} %
    \item[] Justification: Refer to section \ref{app:impact-statement}
    \item[] Guidelines:
    \begin{itemize}
        \item The answer \answerNA{} means that there is no societal impact of the work performed.
        \item If the authors answer \answerNA{} or \answerNo, they should explain why their work has no societal impact or why the paper does not address societal impact.
        \item Examples of negative societal impacts include potential malicious or unintended uses (e.g., disinformation, generating fake profiles, surveillance), fairness considerations (e.g., deployment of technologies that could make decisions that unfairly impact specific groups), privacy considerations, and security considerations.
        \item The conference expects that many papers will be foundational research and not tied to particular applications, let alone deployments. However, if there is a direct path to any negative applications, the authors should point it out. For example, it is legitimate to point out that an improvement in the quality of generative models could be used to generate Deepfakes for disinformation. On the other hand, it is not needed to point out that a generic algorithm for optimizing neural networks could enable people to train models that generate Deepfakes faster.
        \item The authors should consider possible harms that could arise when the technology is being used as intended and functioning correctly, harms that could arise when the technology is being used as intended but gives incorrect results, and harms following from (intentional or unintentional) misuse of the technology.
        \item If there are negative societal impacts, the authors could also discuss possible mitigation strategies (e.g., gated release of models, providing defenses in addition to attacks, mechanisms for monitoring misuse, mechanisms to monitor how a system learns from feedback over time, improving the efficiency and accessibility of ML).
    \end{itemize}
    
\item {\bf Safeguards}
    \item[] Question: Does the paper describe safeguards that have been put in place for responsible release of data or models that have a high risk for misuse (e.g., pre-trained language models, image generators, or scraped datasets)?
    \item[] Answer: \answerNA{} %
    \item[] Justification: All the models and data used in the paper are publicly available.
    \item[] Guidelines:
    \begin{itemize}
        \item The answer \answerNA{} means that the paper poses no such risks.
        \item Released models that have a high risk for misuse or dual-use should be released with necessary safeguards to allow for controlled use of the model, for example by requiring that users adhere to usage guidelines or restrictions to access the model or implementing safety filters. 
        \item Datasets that have been scraped from the Internet could pose safety risks. The authors should describe how they avoided releasing unsafe images.
        \item We recognize that providing effective safeguards is challenging, and many papers do not require this, but we encourage authors to take this into account and make a best faith effort.
    \end{itemize}

\item {\bf Licenses for existing assets}
    \item[] Question: Are the creators or original owners of assets (e.g., code, data, models), used in the paper, properly credited and are the license and terms of use explicitly mentioned and properly respected?
    \item[] Answer: \answerYes{} %
    \item[] Justification: Refer to sections \ref{Introduction}, \ref{related-work} and \ref{sec:experimentation}
    \item[] Guidelines:
    \begin{itemize}
        \item The answer \answerNA{} means that the paper does not use existing assets.
        \item The authors should cite the original paper that produced the code package or dataset.
        \item The authors should state which version of the asset is used and, if possible, include a URL.
        \item The name of the license (e.g., CC-BY 4.0) should be included for each asset.
        \item For scraped data from a particular source (e.g., website), the copyright and terms of service of that source should be provided.
        \item If assets are released, the license, copyright information, and terms of use in the package should be provided. For popular datasets, \url{paperswithcode.com/datasets} has curated licenses for some datasets. Their licensing guide can help determine the license of a dataset.
        \item For existing datasets that are re-packaged, both the original license and the license of the derived asset (if it has changed) should be provided.
        \item If this information is not available online, the authors are encouraged to reach out to the asset's creators.
    \end{itemize}

\item {\bf New assets}
    \item[] Question: Are new assets introduced in the paper well documented and is the documentation provided alongside the assets?
    \item[] Answer: \answerYes{} %
    \item[] Justification: Refer to sections \ref{sec:experimentation}, \ref{app:implementation_details} and \ref{app:dataset_instantiations}
    \item[] Guidelines:
    \begin{itemize}
        \item The answer \answerNA{} means that the paper does not release new assets.
        \item Researchers should communicate the details of the dataset\slash code\slash model as part of their submissions via structured templates. This includes details about training, license, limitations, etc. 
        \item The paper should discuss whether and how consent was obtained from people whose asset is used.
        \item At submission time, remember to anonymize your assets (if applicable). You can either create an anonymized URL or include an anonymized zip file.
    \end{itemize}

\item {\bf Crowdsourcing and research with human subjects}
    \item[] Question: For crowdsourcing experiments and research with human subjects, does the paper include the full text of instructions given to participants and screenshots, if applicable, as well as details about compensation (if any)? 
    \item[] Answer: \answerNA{} %
    \item[] Justification: The paper does not involve crowdsourcing nor research with human subjects
    \item[] Guidelines:
    \begin{itemize}
        \item The answer \answerNA{} means that the paper does not involve crowdsourcing nor research with human subjects.
        \item Including this information in the supplemental material is fine, but if the main contribution of the paper involves human subjects, then as much detail as possible should be included in the main paper. 
        \item According to the NeurIPS Code of Ethics, workers involved in data collection, curation, or other labor should be paid at least the minimum wage in the country of the data collector. 
    \end{itemize}

\item {\bf Institutional review board (IRB) approvals or equivalent for research with human subjects}
    \item[] Question: Does the paper describe potential risks incurred by study participants, whether such risks were disclosed to the subjects, and whether Institutional Review Board (IRB) approvals (or an equivalent approval/review based on the requirements of your country or institution) were obtained?
    \item[] Answer: \answerNA{} %
    \item[] Justification: The paper does not involve any experiments with humans
    \item[] Guidelines:
    \begin{itemize}
        \item The answer \answerNA{} means that the paper does not involve crowdsourcing nor research with human subjects.
        \item Depending on the country in which research is conducted, IRB approval (or equivalent) may be required for any human subjects research. If you obtained IRB approval, you should clearly state this in the paper. 
        \item We recognize that the procedures for this may vary significantly between institutions and locations, and we expect authors to adhere to the NeurIPS Code of Ethics and the guidelines for their institution. 
        \item For initial submissions, do not include any information that would break anonymity (if applicable), such as the institution conducting the review.
    \end{itemize}

\item {\bf Declaration of LLM usage}
    \item[] Question: Does the paper describe the usage of LLMs if it is an important, original, or non-standard component of the core methods in this research? Note that if the LLM is used only for writing, editing, or formatting purposes and does \emph{not} impact the core methodology, scientific rigor, or originality of the research, declaration is not required.
    \item[] Answer: \answerNo{} %
    \item[] Justification: We did not use LLM for any method specific ideas
    \item[] Guidelines:
    \begin{itemize}
        \item The answer \answerNA{} means that the core method development in this research does not involve LLMs as any important, original, or non-standard components.
        \item Please refer to our LLM policy in the NeurIPS handbook for what should or should not be described.
    \end{itemize}
\end{enumerate}

\end{document}